\documentclass{aa}  
\usepackage{graphicx}
\usepackage{txfonts}

\usepackage{xcolor}
\usepackage{hyperref}
\usepackage{multirow}

\newcommand{\oot}{out-of-transit }

\newcommand{\angstrom}{\mbox{\normalfont\AA}\,}
\newcommand{\micron}{um}
\newcommand{\footlink}[1]{\footnote{\href{#1}{#1}}}

\newcommand{\modif}[1]{#1}
\newcommand{\deleted}[1]{\vphantom{#1}}

\newcommand{\modiff}[1]{#1}
\newcommand{\added}[1]{#1}

\begin{document}

   \title{HST/WFC3 transmission spectroscopy of the cold rocky planet TRAPPIST-1h}
   \author{L. J. Garcia \inst{1} \and
           S. E. Moran \inst{2,3} \and
           B. V. Rackham \inst{4, 5} \thanks{51 Pegasi b Fellow} \and
           H. R. Wakeford \inst{6} \and
           M. Gillon \inst{1} \and
           J. de Wit \inst{7} \and
           N. K. Lewis \inst{8}
          }

   \institute{Astrobiology Research Unit, Universit\'{e} de Li\`{e}ge, All\'{e}e du 6 Ao\^{u}t 19C, B-4000 Li\`{e}ge, Belgium \\ \email{lgarcia@uliege.be} \and
              Department of Earth and Planetary Sciences, Johns Hopkins University, 3400 N Charles St, Baltimore, MD 21218, USA \and
              Bay Area Environmental Research Institute/NASA Ames Research Center, Moffett Field, CA 94035, USA \and
              Department of Earth, Atmospheric and Planetary Sciences, Massachusetts Institute of Technology, 77 Massachusetts Avenue, Cambridge, MA 02139, USA \and
              Kavli Institute for Astrophysics and Space Research, Massachusetts Institute of Technology, Cambridge, MA 02139, USA \and
              School of Physics, University of Bristol, HH Wills Physics Laboratory, Tyndall Avenue, Bristol BS8 1TL, UK \and
              Department of Earth, Atmospheric and Planetary Sciences, Massachusetts Institute of Technology, 77 Massachusetts Avenue, Cambridge, MA 02139, USA \and
              Department of Astronomy and Carl Sagan Institute, Cornell University, 122 Sciences Drive, Ithaca, NY 14853, USA
             }

   \date{}
 
  \abstract
   {TRAPPIST-1 is a nearby ultra-cool dwarf star transited by seven rocky planets. We observed three transits of its outermost planet, TRAPPIST-1h, using the G141 grism of the Wide Field Camera 3 instrument aboard the Hubble Space Telescope to place constraints on\protect\deleted{planet h’s} \modif{its potentially cold} atmosphere}
   {In order to deal with the effect of stellar contamination, we model TRAPPIST-1 active regions as portions of a cooler and a hotter photosphere, and generate multi-temperature models that we compare to the out-of-transit spectrum of the star. Using the inferred spot parameters, we produce corrected transmission spectra for planet h under five transit configurations and compare these data to planetary atmospheric transmission models using the forward model \texttt{CHIMERA.}}
   {Our analysis reveals that TRAPPIST-1h is unlikely to host an aerosol-free H/He-dominated atmosphere. While the current data precision limit\modif{s} the constraints we can put on the planetary atmosphere, we find that the likeliest scenario is that of a flat, featureless transmission spectrum in the WFC3/G141 bandpass due to a high mean molecular weight atmosphere ($\geq$1000$\times$ solar), no atmosphere, or an opaque aerosol layer, all in absence of stellar contamination. This work outlines the limitations of modeling active photospheric regions with theoretical stellar spectra, and those brought by our lack of knowledge of the photospheric structure of ultracool dwarf stars. Further characterization of the planetary atmosphere of TRAPPIST-1h would require higher precision measurements over wider wavelengths, which will be possible with \modiff{the James Webb Space Telescope (JWST)}.}
   {}


   \maketitle
\section{Introduction}
While the first detailed atmospheric characterizations of giant exoplanets were carried out during the last two decades (e.g., \citealt{charb2002, deming2013}), the study of smaller planets, that is those comparable to the size of Earth, will become increasingly accessible in the near future. This is especially true for planets orbiting nearby M-dwarfs, offering larger transit signal-to-noise ratios compared to small planets transiting larger stars \citep{vulcans}. This makes the seven planets orbiting TRAPPIST-1, an M8-type dwarf star located 12 parsec away \citep{gillon2017, luger2017}, excellent candidates for detailed atmospheric characterizations with the upcoming James Webb Space Telescope \citep{morley2017, lustig2019}. So far, spectroscopic observations from the Hubble Space Telescope Wide Field Camera 3 (HST/WFC3) have been able to rule out the presence of clear primary hydrogen-dominated atmospheres for all planets \citep{dewit2016, dewit2018, wakeford2019} except for the outermost planet, TRAPPIST-1h, the most likely to have retained such an extended atmosphere \citep{bourrier2017}.
\bigskip\\

The technique of transit spectroscopy, which consists in measuring a planet's transit depths at different wavelengths, remains one of the most successful methods to partially characterize exoplanets atmospheres. However, this method brings with it a major challenge: the effect of stellar contamination (e.g., \citealt{pont2008, apai2018}). Indeed, any spectral difference between the transited chord and the rest of the stellar disk can result in signals of stellar origin able to mimic or hide those of planetary ones \citep{rackham2018, rackham2019}. This concern has led the exoplanetary science community to explore a number of solutions (e.g., \citealt{rackham2019white}), some based on the modeling of active regions probed by the planet during its transit (e.g., \citealt{espinoza2019}), and others addressing the contamination from unocculted active regions (e.g., \citealt{wakeford2019}).
\bigskip\\

In this work, we present the transmission spectrum of TRAPPIST-1h obtained from HST/WFC3 observations in order to rule out the presence of an extended hydrogen-rich atmosphere. We describe our three transit observations and their time-resolved spectra extraction in \autoref{observation}. In \autoref{modeling}, we model the spectroscopic light curves obtained from these measurements, leading to a joint transmission spectrum. We then attempt to model the out-of-transit spectra of TRAPPIST-1 in \autoref{stellar-contamination} in order to deal with the effect of stellar contamination by following the approach of \cite{rackham2018} and \cite{wakeford2019}. By assuming five possible transit configurations, we model the planetary component of the measured spectrum in \autoref{atmosphere}. Finally, we discuss the results of these models and give our conclusions in \autoref{conclusion}.

\section{Observations}\label{observation}
We observed three transits of TRAPPIST-1h with the Hubble Space Telescope (HST) Wide Field Camera 3 (WFC3) as part of HST GO program 15304 (PI: J. de Wit) on UT 2018 July 19 (visit 1), 2019 September 24 (visit 2), and 2020 July 20 (visit \modif{3})\footnote{\modif{These three visits are respectively denoted 03, 02 and 04 in the observing plan}}. Each of the three transits was observed over a five-hour window, each requiring four HST orbits, all consisting in approximately 45 minutes of observation separated by 45 minute gaps due to Earth occultation. We obtained time-series spectroscopy in the 1.12--1.65\,$\mu m$ wavelength band using the G141 grism in scanning mode, spreading the stellar spectrum perpendicularly to its dispersion axis. The scan rate was set to $0\farcs02$\,s$^{-1}$ with an exposure time of 112\,s, resulting in 17-pixel-wide scans acquired in the forward direction only. Each scan is composed of six non-destructive readouts (including the zeroth-read), which we used in our reduction process to remove part of the accumulated background over the complete exposures. A direct image of the target was acquired using
the F139M narrow-band filter at the beginning of each orbit in order to perform G141 wavelength  calibration.

\subsection*{Data reduction}\label{extraction}
We extracted the time-resolved spectra from the three visits following the method presented by \cite{Kreidberg2014} and using the \textsf{prose}\footlink{https://github.com/lgrcia/prose} Python framework (Garcia et al. 2021). \textsf{prose} is a general-purpose tool to build modular astronomical pipelines out of reusable and well-maintained processing blocks. As in  \citet{Kreidberg2014}, we start from the \texttt{ima} pre-calibrated products and process each image through the same steps. As described in \autoref{observation}, the full spectrum in an image is spread along the spatial direction of the detector. During the process, nondestructive readouts are recorded such that a specific subexposure can be constructed by subtracting its readout from the previous one, hence removing part of the accumulated background. We then start by forming subexposures out of nondestructive readouts. For each subexposure:
\begin{enumerate}
  \item we mask bad pixels identified from the \textsf{calcwfc3} pipeline
  \item we compute the wavelength trace using a direct image of the source and the solution from \cite{G141Trace}
  \item we apply a wavelength-dependent flat-field calibration using the method of \cite{G141wvflat}
  \item we interpolate all image rows to the wavelength solution of the direct image so that all values found in a given column correspond to photons from a common wavelength bin
  \item we estimate the subexposure background as the median pixel value within a spectrum-free portion of the image, which we subtract from the subexposure
  \item we cut the spectrum out of the image using a 40-pixel-tall aperture centered in the scanning direction on the spectrum center of light. This large aperture (compared to the 17-pixel-tall spectrum) is manually set in order to include the tail of the WFC3/IR point spread function
  \item we extract a 1D spectrum from this cutout using an optimal-spectrum-extraction algorithm \citep{Horne86}. By using this technique, the fact that more background pixels are contained in our wide aperture does not affect the noise of our extracted signal, as these are optimally weighted. 
\end{enumerate}
Once these seven steps are completed for all subexposures of an image, we interpolate the subexposure spectra to a common wavelength axis and sum them, which yields the final 1D spectrum of the image. Each of the steps described above, applied to an image and its subexposures, are implemented into modular \textsf{prose} blocks, and then assembled into a reduction pipeline made available to the community through the \textsf{prose} Python package.
\bigskip\\

In parallel, we performed a comparative extraction of the  spectra from the three visits using \textsf{iraclis}\footlink{https://github.com/ucl-exoplanets/Iraclis}, an open-source Python package presented by \cite{Tsiaras2016}. Its reduction starts from the \texttt{raw} observation products and goes through the calibration steps of the \textsf{calcwfc3} pipeline\footlink{https://hst-docs.stsci.edu/wfc3dhb/chapter-3-wfc3-data-calibration/3-3-ir-data-calibration-steps}. Then, for every image, the wavelength-dependent photon trajectories along the scanned spectrum are computed, so that fluxes can be extracted within accurately placed polygonal boxes along the wavelength trace (see \citealp[Figure\,6]{Tsiaras2016}). By doing so, the method used in \textsf{iraclis} properly accounts for the off-axis nature of the G141 slitless spectra and refines the wavelength solution of every exposure. 
\bigskip\\

The spectra obtained from \textsf{iraclis} are directly extracted within wavelength bins that we also apply to the \textsf{prose} spectra: 12 bins from 1.1\,$\mu m$ to 1.67\,$\mu m$, leading to 
bins of ${\sim} 0.04$ $\mu m$\ in width. This sampling is chosen to form wavelength bins including complete pixels in the dispersion axis. 
\bigskip\\

\begin{figure*}[htbp!]
    \centering
    \includegraphics[width=\linewidth]{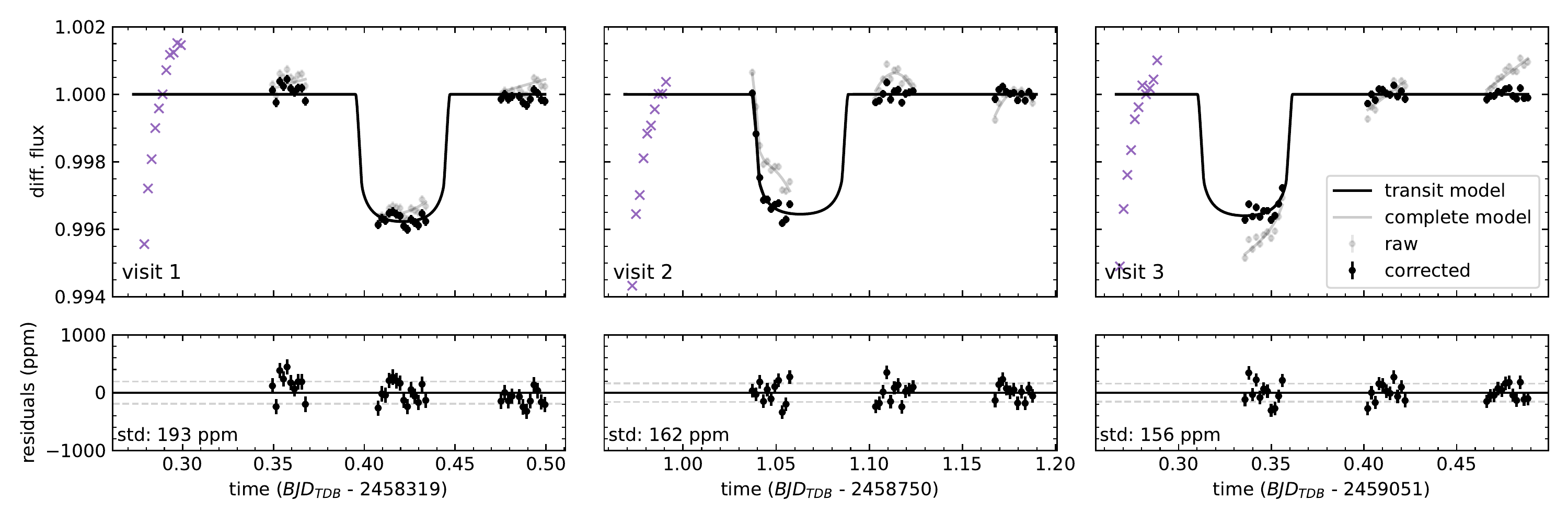}
    \caption{White-light curves of visit \modif{1, 2, and 3} of GO 15304 observations. The first orbit is plotted \protect\deleted{in light grey points but} \modif{as purple crosses and is} not used in the analysis.}
    \label{fig:whites}
  \end{figure*}

The raw data obtained this way feature ramp-like signals characteristic of WFC3 observations. In order to consistently model these signals over all orbits, we discard the first orbit of each visit, which features a higher-amplitude ramp. For the same reason, we discard the first and second exposures of each orbit. The  white-light curves from the three visits are shown in \autoref{fig:whites} (raw data are in light gray) and were obtained by summing the spectra over all wavelength bins.

  
\begin{table}[htbp!]
  \centering
  \begin{tabular}{ccc}
    \hline\hline
    parameter & unit & value \\
    \hline
    $R_{\star}$ & $R_{\odot}$ & $\mathcal{N}(0.1192, 0.0013)$ \\
    $M_{\star}$ & $M_{\odot}$ & \modif{$\mathcal{N}(0.0898, 0.0023)$} \\
    $P$ & $days$ & \modif{$\mathcal{N}(18.7672745, 0.00001876)$} \\
    \modif{$b$} & \modif{-} & \modif{$\mathcal{N}(0.448, 0.054)$} \\
    $T_0$ & BJD tdb & $\mathcal{U}(t_0, t_f)$ \\
  \end{tabular}
  \caption{Priors on TRAPPIST-1 stellar parameters and TRAPPIST-1h orbital parameters. $R_{\star}$ and $M_{\star}$ are respectively the stellar radius and mass \modif{with prior distributions from \citet{agol21}}. $P$ and  \modif{$b$} are TRAPPIST-1h orbital period and  \modif{impact parameters taken from \citet{ducrot2020}}. $T_0$ is  \modif{the} transit mid-time, and the  \modif{transit depth is set to an uninformative uniform prior}. $\mathcal{U}(a, b)$ denotes a uniform distribution bounded by $(a, b)$, and $\mathcal{N}(\mu, \sigma)$ a normal distribution of variance $\sigma^2$ centered on $\mu$. $t_0$ and $t_f$ correspond to the times of the first and last exposures of a given visit.}
  \label{tab:params-priors}
\end{table}

\begin{table}[htbp!]
  \centering
  \begin{tabular}{ccc}
    \hline\hline
    Bandpass ($\mu m$)& $u_1$ & $u_2$\\
    \hline
    1.101 - 1.119 & 0.167 $\pm$ 0.020 & 0.373 $\pm$ 0.029 \\
    1.140 - 1.159 & 0.168 $\pm$ 0.020 & 0.384 $\pm$ 0.030 \\
    1.180 - 1.200 & 0.175 $\pm$ 0.020 & 0.385 $\pm$ 0.029 \\
    1.222 - 1.242 & 0.150 $\pm$ 0.019 & 0.362 $\pm$ 0.028 \\
    1.265 - 1.286 & 0.146 $\pm$ 0.018 & 0.338 $\pm$ 0.026 \\
    1.310 - 1.332 & 0.177 $\pm$ 0.018 & 0.352 $\pm$ 0.026 \\
    1.355 - 1.379 & 0.230 $\pm$ 0.019 & 0.410 $\pm$ 0.028 \\
    1.403 - 1.428 & 0.326 $\pm$ 0.014 & 0.359 $\pm$ 0.021 \\
    1.453 - 1.479 & 0.289 $\pm$ 0.016 & 0.378 $\pm$ 0.023 \\
    1.504 - 1.530 & 0.224 $\pm$ 0.019 & 0.396 $\pm$ 0.027 \\
    1.558 - 1.585 & 0.178 $\pm$ 0.019 & 0.373 $\pm$ 0.027 \\
    1.612 - 1.641 & 0.135 $\pm$ 0.018 & 0.331 $\pm$ 0.026 \\

  \end{tabular}
  \caption{Quadratic limb darkening parameters obtained from ExoCTK \citep{exoctk}.}
  \label{tab:ldc}
\end{table}

\section{Light-curve modeling}\label{modeling}

In the following sections, we present our modeling of the transmission spectra obtained from \textsf{iraclis} and \textsf{prose} and validate our results based on their comparison.

\subsection{Systematic models comparison}\label{model-com}
Spectroscopic light curves obtained with HST display a high level of wavelength-, visit-, and orbit-dependent systematic signals (\citealt{wakeford2016} and references therein). Similarly to previous studies, we model these signals using empirical functions of time $t$ and HST orbital phase $\phi,$ which can take a variety of forms. For this study, we employ polynomials such that the white-light curve of a given visit can be modeled as:
\begin{equation} \label{model}
F(t, \phi) = T(t) + P(t, \phi) + \epsilon(t),
\end{equation}
where $T$ is the transit light curve model, $P$ is a polynomial systematic model of time $t$ and HST orbital phase $\phi$, and $\epsilon$ is a Gaussian noise vector. As in \citet{wakeford2016}, we selected the combination of polynomial orders of $t$ and $\phi$ through a model comparison based on the minimization of the Akaike information criterion (AIC, \citealt{aic}), such that our models are predictive while being limited in the number of parameters included. By allowing polynomials of up to order three and excluding cross terms, we end up with a set of 16 models to compare.

For a given visit and for each order combination, we infer the best transit and systematic model parameters in a Bayesian framework using \textsf{exoplanet} \citep{exoplanet}\footlink{https://docs.exoplanet.codes/en/latest/}, making use of the inference framework \textsf{PyMC3}\footlink{https://docs.pymc.io/} \citep{pymc3} and the transit light curve model from \citet{starry}. The priors we use for the orbital parameters of TRAPPIST-1h are listed in \autoref{tab:params-priors}, with $T_0$ having a uniform prior over the full time of the visit. This uninformative prior on the mid-transit time is used to encompass the wide transit time variations (TTVs) observed for the TRAPPIST-1 planets, reaching more than 1 hour for TRAPPIST-1h \citep[Figure~2]{agol21}. Finally, we use a quadratic limb-darkening model with parameters fixed to the values in \autoref{tab:ldc}, obtained from ExoCTK\footlink{https://exoctk.stsci.edu/limb\_darkening} \citep{exoctk} using the PHOENIX ACES stellar atmosphere model \citep{husser2013}. \deleted{ Thus, all transit model parameters are fixed except for the transit mid-time, planetary, and stellar radii.}

Given our data, we find the maximum a posteriori model parameters using the BFGS algorithm \citep{bfgs}, as implemented in \textsf{scipy}\footlink{https://docs.scipy.org/doc/scipy/reference/optimize.minimize-bfgs.html}, and use this maximized posterior to compute the AIC for each order combination. Finally, for each visit, we retain the model with the minimal AIC (\autoref{fig:aics}). While the best model is found by modeling the white-light curve of each visit, we also use it to model the systematic errors for the spectroscopic light curves, keeping the same form but allowing its parameters to vary from one wavelength bin to another.

\subsection{Transmission spectra inference}

We model the spectroscopic light curves from each visit using the systematic models found in \autoref{model-com}, now using the wavelength-dependent expression of the flux:
\begin{equation}
  F_{\lambda}(t, \phi) = T_{\lambda}(t) + P_{\lambda}(t, \phi) + \epsilon_{\lambda}(t)
,\end{equation}
and following the same notation as in \autoref{model} with $\lambda$ denoting the wavelength bin. As we are interested in the wavelength-dependent transit depth of TRAPPIST-1h, we set an uninformative prior on its wavelength-dependent radius:
\begin{equation}
  R_{p, \lambda} \sim \mathcal{U} (0.5, 2) R_{\oplus}
,\end{equation}
where $\mathcal{U}(a, b)$ is a uniform distribution bounded by $a$ and $b$. Again, we use the orbital parameters and priors listed in \autoref{tab:params-priors}. As in \autoref{model-com}, we adopt a quadratic limb-darkening model, the coefficients of which are held fixed to the values found in \autoref{tab:ldc}. For each visit, we estimate the wavelength-dependent transit depths and their uncertainties through an Hamiltonian Monte Carlo analysis of the data, making use of the \textsf{exoplanet} package \citep{exoplanet}.

\begin{figure}
  \centering
  \includegraphics[width=\linewidth]{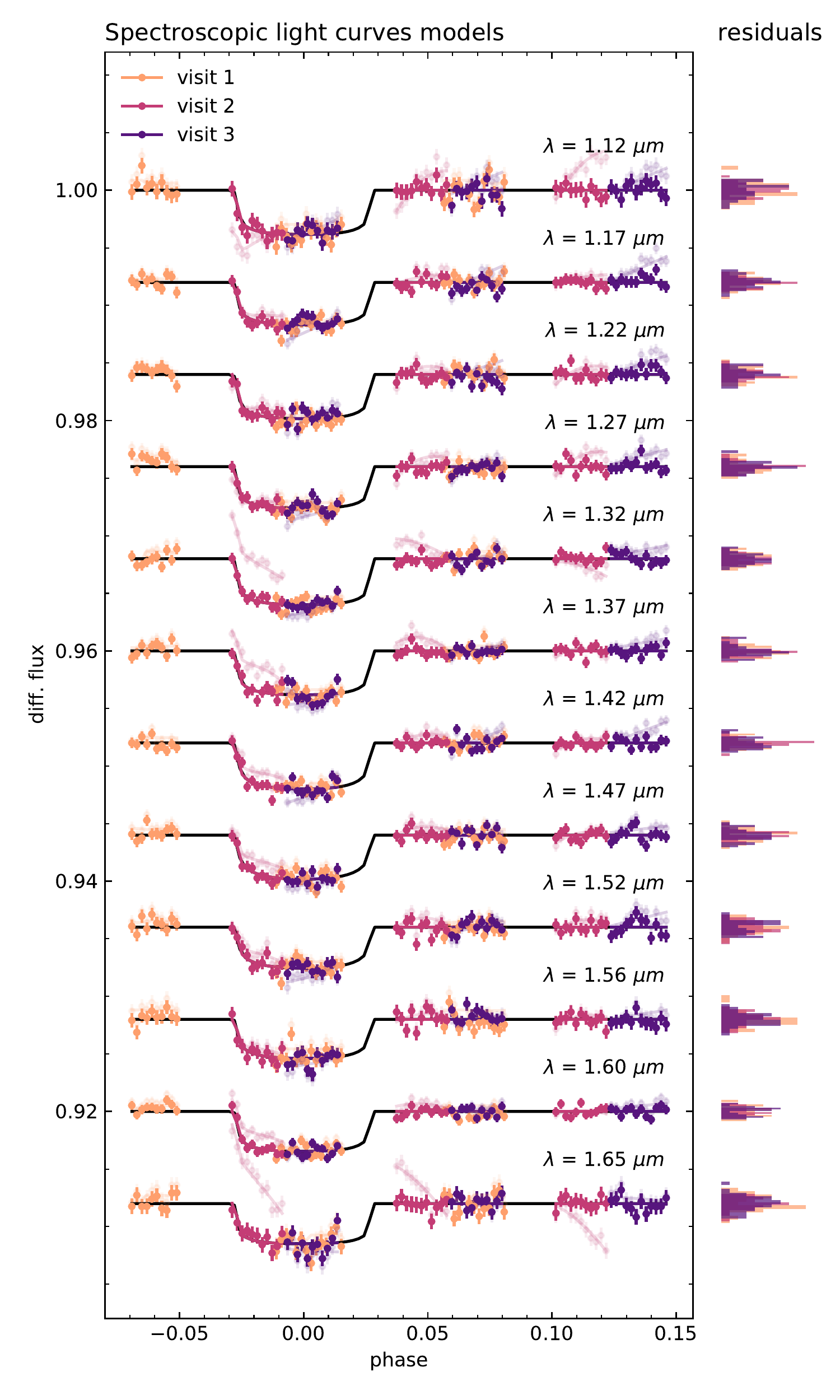}
  \caption{Joint spectroscopic light-curve model for the three visits\modiff{, each plotted with a single color}. The raw data are represented by \modiff{transparent points in the backgorund}, and systematic models \modiff{overalaid as transparent lines}. Solid points correspond to light curves corrected for the systematic error models, and solid black lines correspond to the inferred transit models per wavelength band. Finally, histograms of the residuals between the data and models are plotted to the right of each spectroscopic light curve.
  \label{fig:spectroscopic}
  }
\end{figure}

\begin{figure*}[htbp!]
    \centering
    \includegraphics[width=0.8\linewidth]{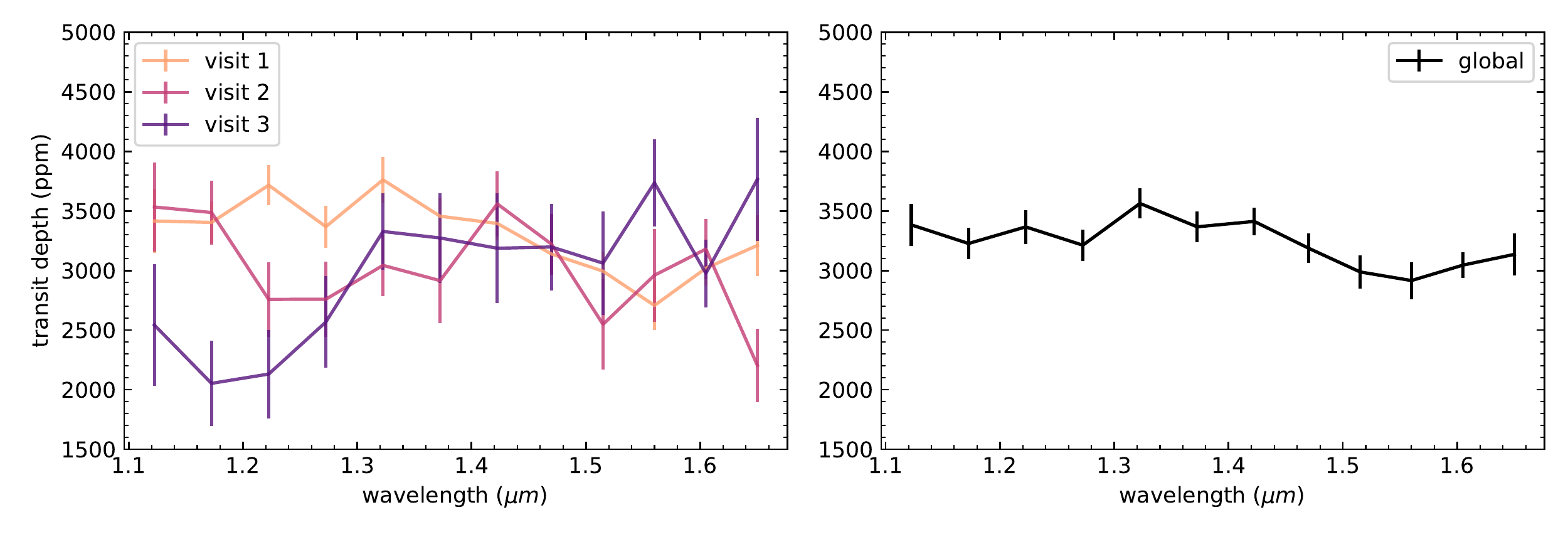}
    \caption{Transmission spectra obtained for each visit individually (left) as well as the one obtained through a joint analysis (right).}
    \label{fig:transp}
\end{figure*}

Once this is done on individual visits (see the transmission spectra in \autoref{fig:transp}, left plot), we proceed to a global analysis of all visits using the priors listed in \autoref{tab:params-priors}, and derive the global transmission spectrum shown in the right plot of \autoref{fig:transp} with the spectroscopic light curves shown in \autoref{fig:spectroscopic}. We do that by allowing each visit to have its own systematic model, held to the ones found in the previous section (shown in \autoref{tab:t0syst}) with free parameters. We verify that allowing for nonperiodic transits while accounting for transit time variations leads to the same transmission spectrum. Hence, we use the results obtained in the case of strictly periodic transits in the reminder or our analysis. The transit times obtained by considering each visit individually are reported in \autoref{tab:t0syst}.

\begin{table}
  \centering
  \begin{tabular}{ccc}
    \hline\hline
    visit & transit time & systematics model \\
    \hline

    \modif{1} & \modif{2458319.4206 $\pm$ 0.0030} & \modif{$1 + \phi$} \\
    \modif{2} & \modif{2458751.0660 $\pm$ 0.0003} & \modif{$1 + \theta + \phi^2$} \\
    \modif{3} & 2459051.3365 $\pm$ 0.0008 & $1 + \theta + \phi$ \\
    
  \end{tabular}
  \caption{Transit times inferred individually from each visit as well as the selected systematic error model}
  \label{tab:t0syst}
\end{table}

Finally, we check that the fluxes obtained from \textsf{prose} and \textsf{iraclis} lead to similar transmission spectra. We also compare these transmission spectra to spectra obtained from an independent extraction, masking, and modeling following the methodology presented in \cite{wakeford2019}. The inferred transmission spectra are consistent across the three analyses  (\autoref{fig:iracomp}).
\autoref{fig:transp} highlights variations of the transmission spectra from one visit to another, which seem more pronounced at shorter wavelengths. To study the possible astrophysical origin of these variations, we proceed to the assessment of stellar contamination.


\section{Stellar contamination}\label{stellar-contamination}
\subsection{Photosphere model}

To properly interpret the global transmission spectrum inferred in \autoref{modeling}, and to understand its variations between visits, we need to carefully assess the possible impact of  stellar contamination on our transit observations.
We adopt the same approach as \citet{wakeford2019} by modeling the median \oot spectrum of TRAPPIST-1, accounting for surface heterogeneities such as spots and faculae. Having a model of the stellar disk alone then allows us to correct the measured transmission spectrum from the stellar one, given the portion of the photosphere that is occulted. As in previous studies (e.g., \citealt{rackham2018, rackham2019}), we assume that the complete photosphere can be modeled as a combination of portions of cooler and hotter areas so that the spectrum of the star can be expressed as:

\begin{equation}\label{eq:phot-model}
F = (1 - s - f)F_{phot} + sF_{spot} + fF_{fac},
\end{equation}
where $s$ and $f$ are the covering fractions of spots and faculae (here simply denoting the cooler and hotter portions of the photosphere) and $F_{phot}$, $F_{spot}$, and $F_{fac}$ are the intrinsic spectra of the quiescent photosphere, spots, and faculae, respectively. This simple model assumes that surface heterogeneities have spectra similar to that of a global photosphere of a different temperature. We note that covering fractions $f$ and $s$ can be null, meaning that the model reduces to only one or two temperature components. To model the individual components of the TRAPPIST-1 photosphere, we use the BT-Settl stellar atmosphere models \citep{allard2012}, which are specifically suited for cool stars, down to 1500K. Given that our observations are conducted in the infrared, these are more sensitive to cooler components, which justifies the choice of a model available at cooler temperatures.

\subsection{Models and data preparation}
Fitting \autoref{eq:phot-model} to our data requires some preliminary steps. Indeed, BT-Settl theoretical spectra are provided on a grid of stellar metallicities, log-gravity, and effective temperatures. From this grid, we only keep spectra corresponding to $[Fe/H]=0$, which is close to the inferred value of $[Fe/H]=0.04$ \citep{grootel2018}. We linearly interpolate the models along the log-gravity parameter in order to produce spectra with $\log_{10}(g)= 5.22$ \citep{grootel2018}. By fixing these values, only the effective temperature of the models will be varied during their inference (except for the case discussed in \autoref{fitoot}), ranging from 1500K to 5000K in steps of 100K. We then convert these models to flux density at Earth, accounting for TRAPPIST-1's angular diameter, which is calculated as, $\alpha=(R_\star/d)^2$ using the stellar radius $R_\star = 0.119\,R_\odot$ \citep{agol21} and the distance $d=40.54$\,\modif{ly} \citep[DR2]{gaia}. Finally, we convolve the models with the WFC3 point spread function\footlink{https://hst-docs.stsci.edu/wfc3ihb/chapter-7-ir-imaging-with-wfc3/7-6-ir-optical-performance} and sample them according to the WFC3 IR detector (50$\angstrom$ pixels).

\begin{figure}[htbp!]
  \centering
  \includegraphics[width=1.\linewidth]{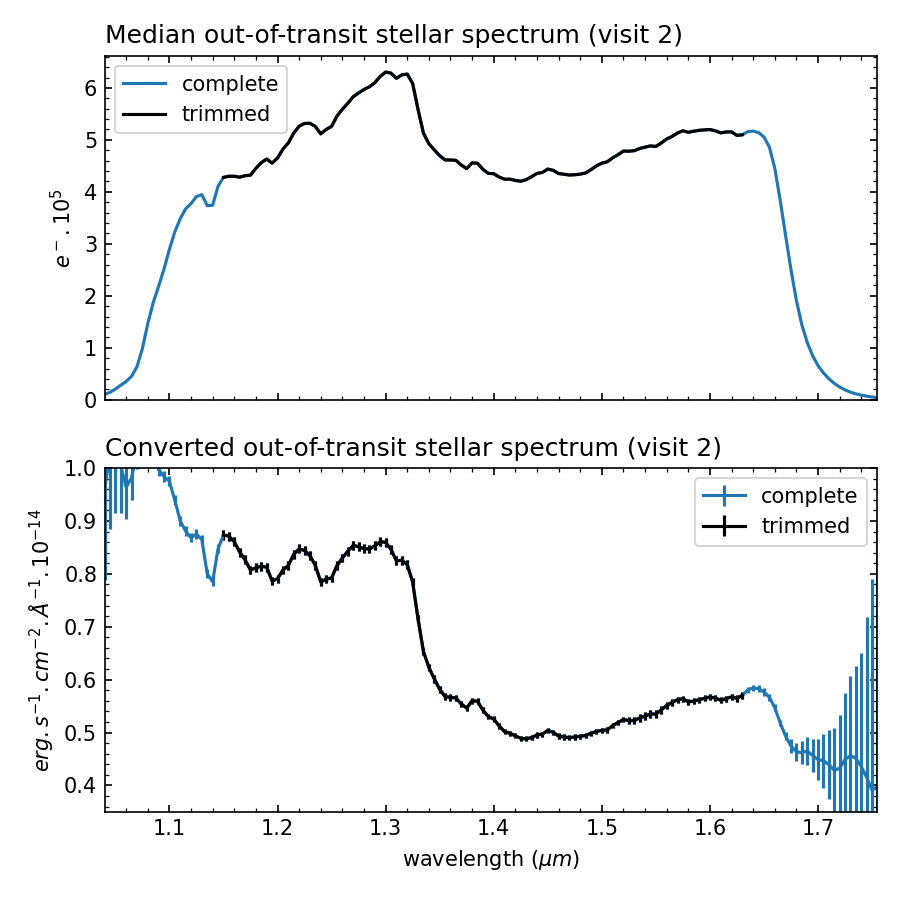}
  \caption{Conver\modif{s}ion of visit 2 \oot stellar spectrum. The trimmed portion is used to perform the model fit.}
  \label{fig:conv-oot}
\end{figure}

The dataset we model consists of three \oot spectra of TRAPPIST-1, one for each visit, and we fit each of them to a theoretical model. For each visit, we build a median spectrum over time, from which we correct the systematic signal that strongly affects the variations of its mean value (as modeled in \autoref{model-com} and shown in \autoref{fig:spectroscopic}). To be compared with the models, we convert these spectra from $e^{-1}\,\mathrm{s}^{-1}\,\angstrom^{-1}$ to $\mathrm{erg}\,\mathrm{s}^{-1}\,\mathrm{cm}^{-2}\,\angstrom^{-1}$ by correcting for the WFC3 IR detector sensitivity\footlink{http://www.stsci.edu/hst/instrumentation/wfc3/documentation/grism-resources/wfc3-g141-calibrations}, accounting for exposure time (112 s) and wavelength passband. WFC3 sensitivity quickly drops to zero outside of the band it covers, meaning that correcting for it (through division) leads to large errors on the tails of our spectra. While we could account for it in the fitting procedure, we find a better convergence by trimming the measured spectra to the 1.15--1.63\,$\micron$ band (see \autoref{fig:conv-oot}).\\

Finally, we add to our dataset the Pan-STARSS g, r, i, and z photometry of TRAPPIST-1 \citep{panstars}. These measurements were added to constrain the visible portion of the stellar spectrum, a choice motivated by the discrepancy observed when comparing the multi-component model from \citealt{wakeford2019} to these data (see \autoref{fig:hw}).

\subsection{Fitting the \oot spectra}\label{fitoot}
\begin{figure*}[htbp!]
  \centering
  \includegraphics[width=1.\linewidth]{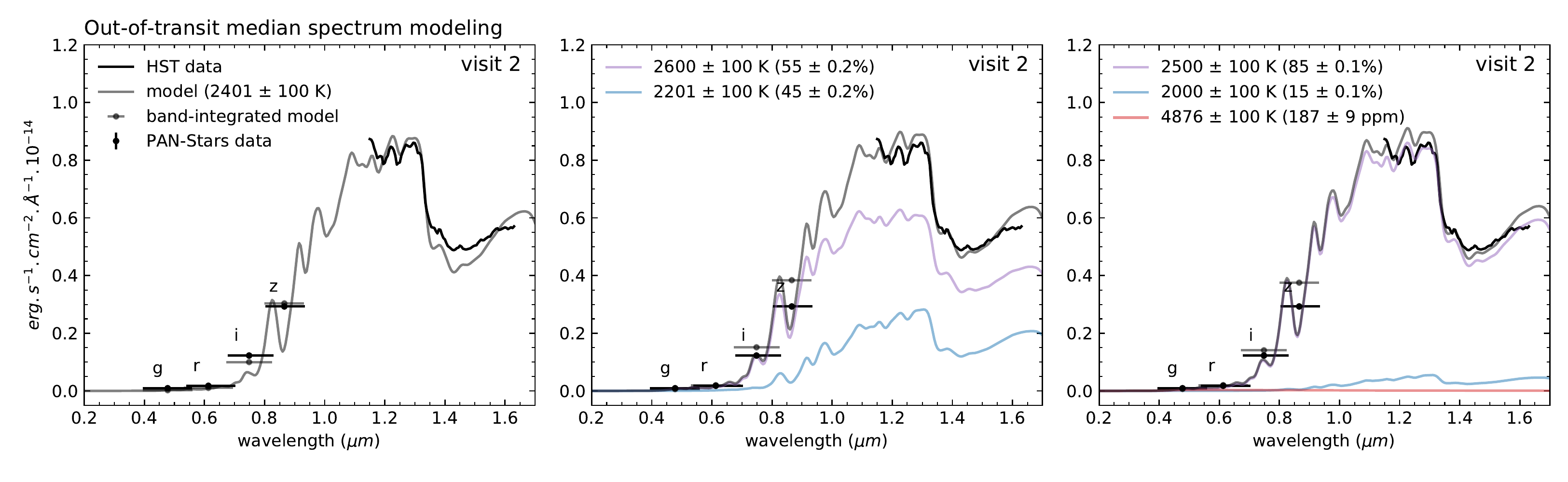}
  \caption{\textit{1T}, \textit{2T}, and \textit{3T} models of the visit 2 \oot stellar spectrum.}
  \label{fig:oot_fit}
\end{figure*}

We fit the dataset previously described to \autoref{eq:phot-model} for each visit individually, accounting for three distinct cases: a quiescent photosphere (1T), a photosphere with either spots or faculae (2T), and a photosphere with both spots and faculae (3T). 
For each visit and for each model, the best maximum a posteriori model parameters are found by following a brute force approach. This procedure consists of estimating the likelihood of the model in a grid of discrete temperature combinations with 100\,K steps (see \autoref{tab:grids_details}). For each point in the grid, that is, for each temperature combination, the best covering fractions (between 0 and 1) are found using the BFGS optimization algorithm. This leads to a complete sampling of the likelihood function over all possible temperatures, and allows for exploration of the multi-modal nature of the likelihood distribution over the component temperatures. In this grid, the maximum likelihood parameters are retained, and refined with a parallelized Markov Chain Monte Carlo using \textit{emcee} \citep{emcee}\modif{, this time in a continuous parameter space,} from which we estimate their uncertainties (see \autoref{oot-corner}). We find that the 2T model likelihood has a well-defined maximum (\autoref{fig:2T-ll}) while the 3T model is bimodal (see \autoref{fig:3T-ll}). \autoref{fig:oot_fit} shows the fitted out-of-transit spectrum for models 1T, 2T, and 3T, obtained by following this procedure on visit \modif{1} data. \modif{As the theoretical stellar spectra are produced in 100K steps, we use this value as the uncertainty on the temperatures of the inferred components  (see e.g., the temperature distributions in \autoref{fig:3T-ll})}

\begin{figure*}[htbp!]
  \centering
  \includegraphics[width=.9\linewidth]{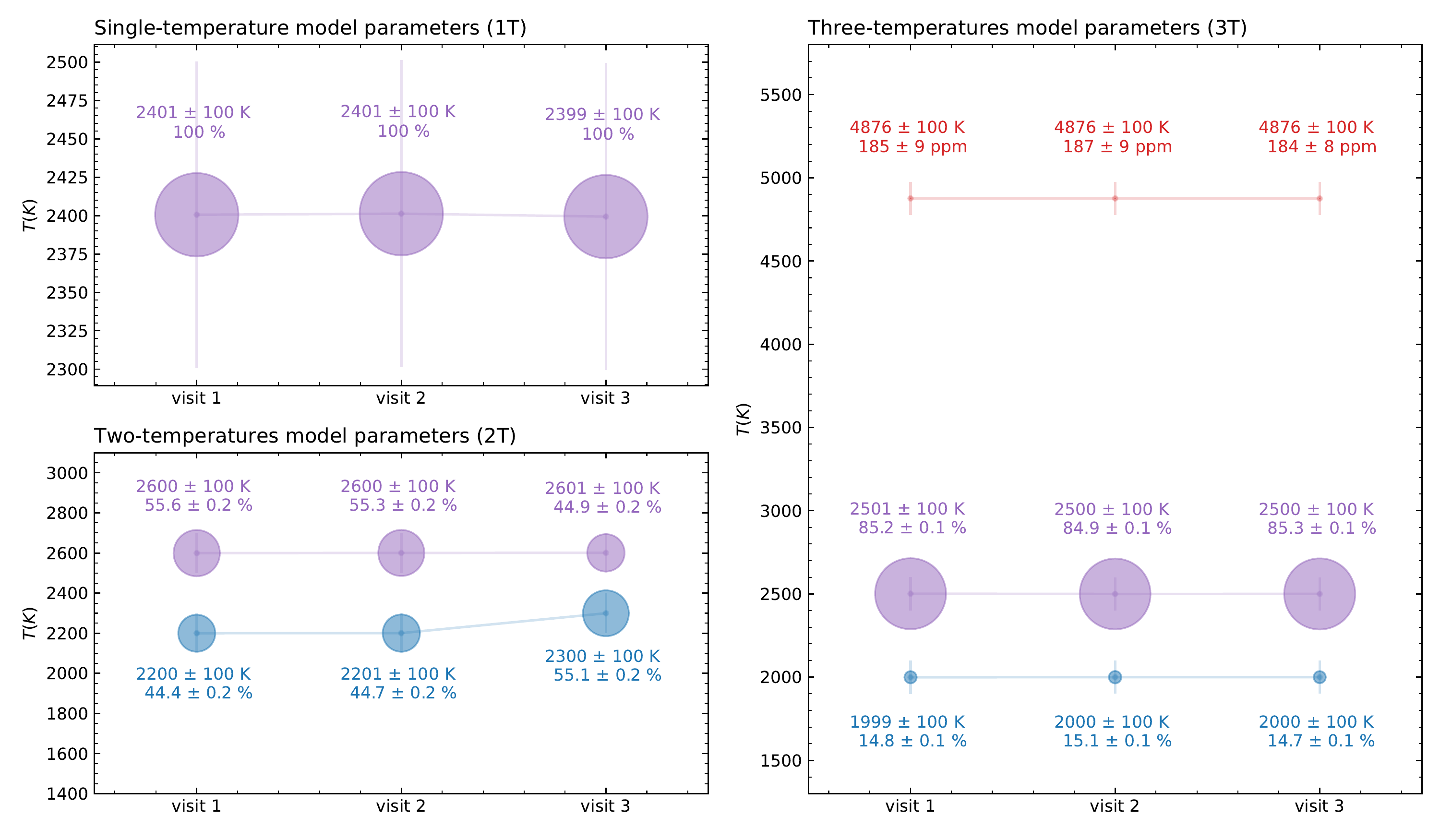}
  \caption{Inferred temperatures and covering fraction for the \textit{1T}, \textit{2T}, and \textit{3T} models on each visit.}
  \label{fig:phot-cons}
\end{figure*}

As in \citet{wakeford2019}, chi-squared statistics indicate that none of our models provide a good fit to the data from any visit. Moreover, the \textit{1T} model shows a larger discrepancy ---which was also observed by \citet{wakeford2019}--- when modeling the HST/WFC3 TRAPPIST-1g \oot spectrum against PHOENIX ACES models \citep{husser2013}. In this latter study, this discrepancy is associated to a poor constraint on the stellar radius $R_{\star}$. This led the authors to introduce a scaling parameter to the stellar radius with the overall motivation being that it would account for  the underestimation of the radii
of low-mass stars by theoretical models. However, introducing such a factor not only would affect $\alpha$ but also the log surface gravity $\log_{10}(g) = GM_{\star}/R_{\star}^2$ involved in the stellar model parameters. Using such a radius scaling, varying the value of $\alpha$ and the log surface gravity provides a better fit to the data (see \autoref{fig:scaled_oot_fit}) but leads to $R_\star = 0.1134 \,\pm\, 0.0013 \,  R_{\odot}$, which is different from the radius inferred by \citet{agol21} by  ${>}3\sigma$
. We decided not to consider this scaling in our study (we refer the interested reader to the analysis of \citet{wakeford2019}) and instead use the unscaled radius found in \citet{agol21}.

In order to understand the differences in the spectra obtained from the three visits, we plot the inferred temperatures and covering fractions found for each of them in \autoref{fig:phot-cons}. Using the transmission spectrum correction described in the following section, we find that none of these models are able to explain the spectroscopic variability we observe. However, we note that only visit \modif{1} contains the bottom of the transit signal and results in a transmission spectrum very similar to the one obtained through a joint analysis of all three visits. For this reason, we assume that the observed variability is not of astrophysical origin but is instead due to the partial coverage of two of the transits observed, combined with \deleted{correlated noise} \modif{the systematic effects} due to the instrument \modif{(see \autoref{fig:corre})}. With this conclusion, we assume that the photospheric structure of TRAPPIST-1 is consistent over the three visits and compare model \autoref{eq:phot-model} to the median visits-combined \oot spectrum, leading to the temperature components reported in \autoref{tab:configurations} for the 1T, 2T, and 3T models. We notice that the hotter component of model 3T ($\sim$5000K) corresponds to the maximum effective temperature in our model's parameter space. Hence, this inferred temperature should be considered as a lower limit to the hot-spot temperature. 

We note that the inferred 3T model, which includes bright spots covering a very small fraction of the stellar disk, is compatible (in order of magnitude) with the findings of \citet{morris2018}. In this latter study, the authors found that 32 ppm of $5300 \pm 200$\,K bright spots are able to explain why the 3.3-day variability observed in the TRAPPIST-1 K2 light curve is undetectable in the \textit{Spitzer} 4.5\,$\micron$ band.

In order to compare our analysis to that of \cite{wakeford2019}, we repeat the modeling procedure described in this section using PHOENIX ACES models, which provides a poorer fit to the data. While the inferred stellar component temperatures are different from the ones found using the PHOENIX BT-Settl models, a bimodal likelihood distribution is also observed with these alternative models. For this reason, we base our study on the results presented in this section, which were obtained using the PHOENIX BT-Settle theoretical spectra.

\subsection{Corrected transit depths}

\begin{figure}[htbp!]
  \centering
  \includegraphics[width=\linewidth]{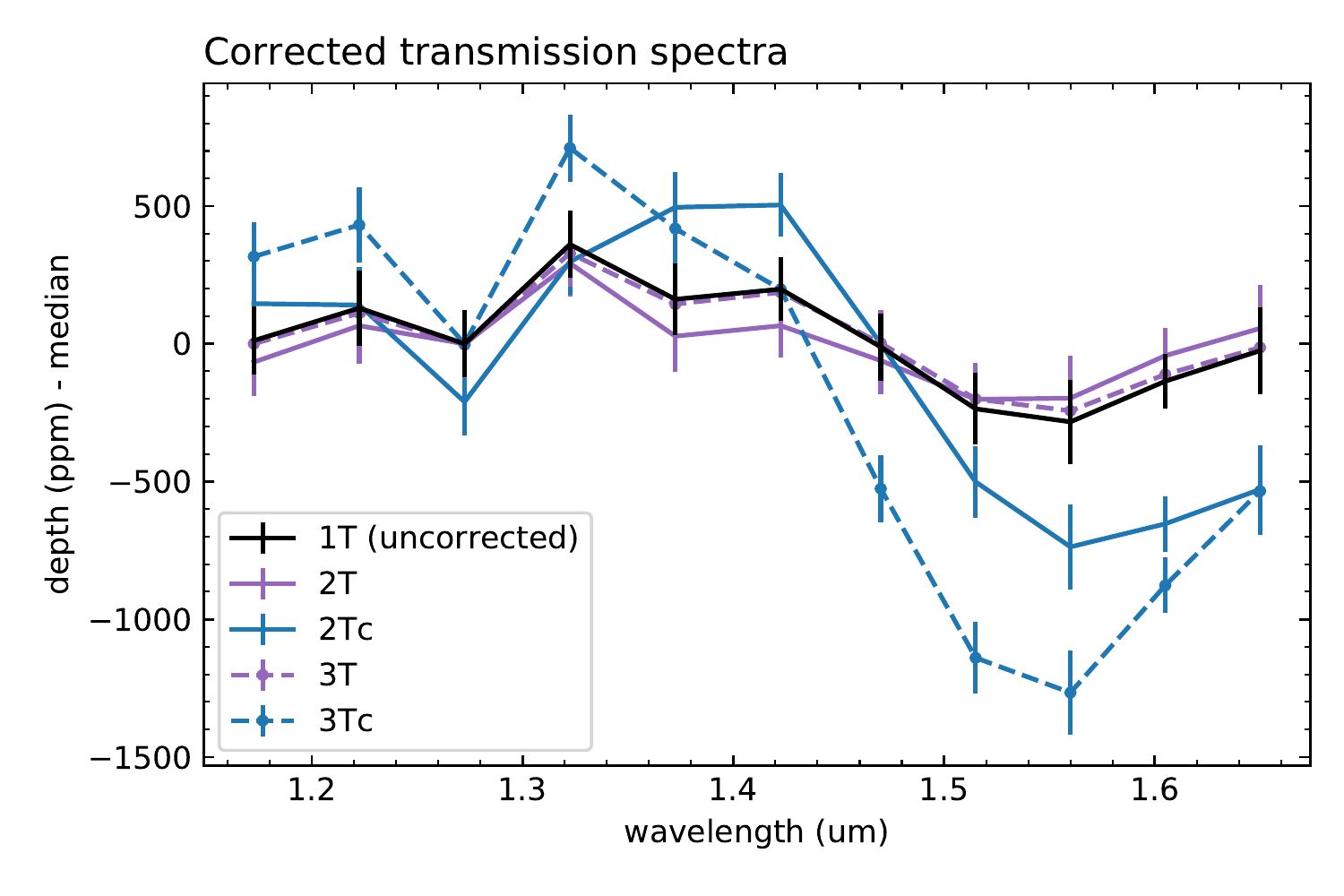}
  \caption{Corrected transmission spectra under the five configurations reported in \autoref{tab:configurations}. Colored solid lines correspond to the two-temperature models (2T) and dotted lines denote the three-temperature models (3T). Purple lines represent spectra of planet h transiting the quiescent component of the stellar photosphere (2T and 3T) while blue lines indicate spectra of planet h transiting the cooler component of the stellar photosphere (2Tc and 3Tc).}
  \label{fig:corr_trans}
\end{figure}

Using the models found in the previous section, the measured transmission spectra can now be corrected for the effect of stellar contamination. We apply corrections for five different configurations, corresponding to the 1T, 2T, and 3T models with the transit chords either passing completely over the base ($F_{phot}$) or the cooler ($F_{spot}$) portion of the photosphere. Due to its very small coverage, we do not consider a configuration where only the hotter component ($F_{fac}$) is occulted. Then, for each configuration and  wavelength bin, the corrected transit depth can be expressed as:
\begin{equation}
\delta_c = \delta \times \frac{F}{F_{occ}},
\end{equation}
where $\delta$ is the measured transit depth,  $F_{occ}$ is the flux of the stellar disk occulted by the planet over the complete transit chord, and $F$ is the overall flux of the stellar disk. As in \citet{wakeford2019}, we note that $F_{occ}$ could be a linear combination of the flux components used to model the stellar disk. However, we decide to explore only the extreme cases where the occulted portion of the photosphere is made of a single temperature component. Applying this model to the five configurations outlined in \autoref{tab:configurations} leads to the corrected transmission spectra plotted in \autoref{fig:corr_trans}. Under the assumptions described above, these spectra should be of planetary origin, with the contribution from the star being modeled and corrected out. We denote the two-component and three-component photosphere models  2Tm and 3Tm, with the transit chord being over the 2600K and 2500K component,  respectively. On the other hand, 2Tc and 3Tc correspond to transit configurations where the chord is over the cooler component of 2200K and 2000K, respectively. Finally, 1T denotes a quiescent photosphere at 2400K for which no spectrum correction is required.

\begin{table}
  \centering
  \begin{tabular}{ccc}
    \hline\hline
    model & components & min-max (K)\\
    \hline
    1T & $F_{phot}$ & 1500-5000\\
    \hline
    \multirow{2}{*}{2T} & $F_{spot}$ & 1500-5000\\
    & $F_{phot}$ & 1500-5000\\
    \hline
    \multirow{3}{*}{3T} & $F_{spot}$ & 1500-2300\\
    & $F_{phot}$ & 2300-2700\\
    & $F_{fac}$ & 2700-5000\\

  \end{tabular}
  \caption{Temperature grids where covering fractions and model likelihoods are estimated, sampled in steps of 100K. The covering fractions for all components are allowed to vary from 0 to 1.}
  \label{tab:grids_details}
\end{table}

\begin{table}
  \centering
  \begin{tabular}{cccc}
    \hline\hline
    model & covering fractions & temperatures  & configuration\\
    \hline
    1T & 100\% & 2400 $\pm$ 100 K & 1T\\

    \hline
    
    \multirow{2}{*}{2T} & 55.5 $\pm$ 0.1 \% & 2600 $\pm$ 100 K & 2Tm \\
    & 44.5 $\pm$ 0.1 \% & 2200 $\pm$ 100 K & 2Tc \\
    
    \hline
    
    \multirow{3}{*}{3T} & 85.1 $\pm$ 0.1 \% & 2500 $\pm$ 100 K & 3Tm\\
    & 14.9 $\pm$ 0.1 \% & 2000 $\pm$ 100 K & 3Tc\\
    & 185.2 $\pm$ 8.8 ppm & 4876 $\pm$ 100 K & 3Th\\

  \end{tabular}
  \caption{Multiple-temperature model parameters inferred from the visits-combined \oot spectrum. For each of the \textit{T1}, \textit{T2}, and \textit{T3} models, we consider a configuration where the stellar disk is occulted across a unique temperature component. The name of the configuration associated to each component is reported in the last column.}
  \label{tab:configurations}
\end{table}

\section{Planetary spectrum}
\label{atmosphere}

In this section, we use the plausible stellar photospheric models from the previous sections to guide our planetary atmospheric models and interpret the planetary transmission spectra. Our primary aim here is to use the potential planetary and stellar scenarios together to find the most likely scenario for each. In so doing, we (1) rule out possible end-member photospheric configurations in cases where they require transmission spectra without a plausible atmospheric model and (2) rule out potential planetary atmospheric conditions when none of the possible transmission spectra (given the allowed  photospheric configurations) are well fitted by an atmospheric model. We summarize the results of the allowable stellar photospheric configurations given the planetary models in \autoref{tab:config_planet} and the \modiff{possible} planetary models given the data in \autoref{tab:planetatmos}.

\subsection{Transmission forward models}
Once we had generated all transit spectrum scenarios (1T, 2Tc, 2Tm, 3Tc, 3Tm, and 3Th), we computed model atmospheric spectra for TRAPPIST-1 h against which we were able to compare the data. We did not run models to compare against scenario 3Th, as the hot spot component is too small to be entirely transited by the planet. As in \cite{wakeford2019}, we used the forward modeling capabilities of the one-dimensional radiative transfer code \texttt{CHIMERA} \citep{line2013}, as has previously been modified for the inner TRAPPIST-1 planets \citep{batalha2018,moran2018}. \texttt{CHIMERA} uses the correlated-\textit{k} method for radiative transfer and the five-parameter, double, gray, one-dimensional temperature-pressure profile of \cite{guillot2010}. We run models in chemical equilibrium, which draw from a pre-computed grid at each temperature and metallicity bin, where we consider a temperature of 170 K and metallicities of between 1  and 1000 times solar. While we include Rayleigh scattering due to H$_2$/He, we neglect any furthering scattering or absorption from clouds or hazes explicitly. Given the precision of our observations, we always first consider the simplest aerosol-free atmospheric models in exploring possible atmospheric signatures. For each model atmosphere, we use a planetary mass of 0.755 M$_{\oplus}$ and a solid body radius of 0.326 R$_{\oplus}$, as determined by \cite{agol21}. 

Because the planetary equilibrium temperature is very cold, namely $\sim$170 K, the atmospheric scale height \textit{H} even for a solar metallicity (2.32 $\mu$) atmosphere is quite small despite the low gravity of the planet (5.6 ms$^{-2}$, \citealt{agol21}), with \textit{H}$\sim$100 km. For reference, \textit{H} in an N$_2$ atmosphere (28 $\mu$) for TRAPPIST-1~h would be $\sim$9 km. For our solar-metallicity model atmosphere, we include opacities from water, methane, carbon monoxide, carbon dioxide, ammonia, molecular nitrogen, and H$_2$/He collision-induced-absorption (CIA) \citep{freedman2008,freedman2014}. For the water- and methane-rich  ten-times-solar atmosphere, we include only water, methane, and H$_2$-CIA in order to produce the largest molecular features in the WFC3 G141 bandpass. These larger features result from the water and methane abundance increasing before the mean molecular weight increases enough to sufficiently shrink the atmospheric scale height \citep{kempton2017,moran2018}. Given the relatively flat transit spectrum scenarios of the data, we explicitly model this water- and methane-rich ten-times-solar atmosphere because it is the least likely model to fit the observed data; though we note that for most planetary transit depths, this is still rejected at less than 3$\sigma$. Additionally, we model a CO$_2$-rich atmosphere at 200 times solar, including opacities from CO$_2$, CH$_4$, and CO. 

Finally, we also model a pure N$_2$ atmosphere without Rayleigh scattering, which in the near-infrared (NIR) region of WFC3 G141 is also representative of an airless body, as well as potentially representing a heavily aerosol-laden atmosphere. \citet{moran2018} showed that, for the inner TRAPPIST-1 planets, aerosols cannot produce a flat line transmission spectrum in the HST WFC3 G141 band in a low mean molecular weight atmosphere. However, the precision of the WFC3 G141 data combined with TRAPPIST-1h's small scale height is such that we cannot rule out aerosols as a potential fit to the data here. Though this ``flat line'' scenario fits the data best (i.e., it produced the lowest reduced-$\chi^2$ value), we cannot completely discount any of the modeled spectra to high significance given the data.

\subsection{Model results with HST}
\begin{figure}[htpb!]
    \centering
    \includegraphics[width=0.9\linewidth]{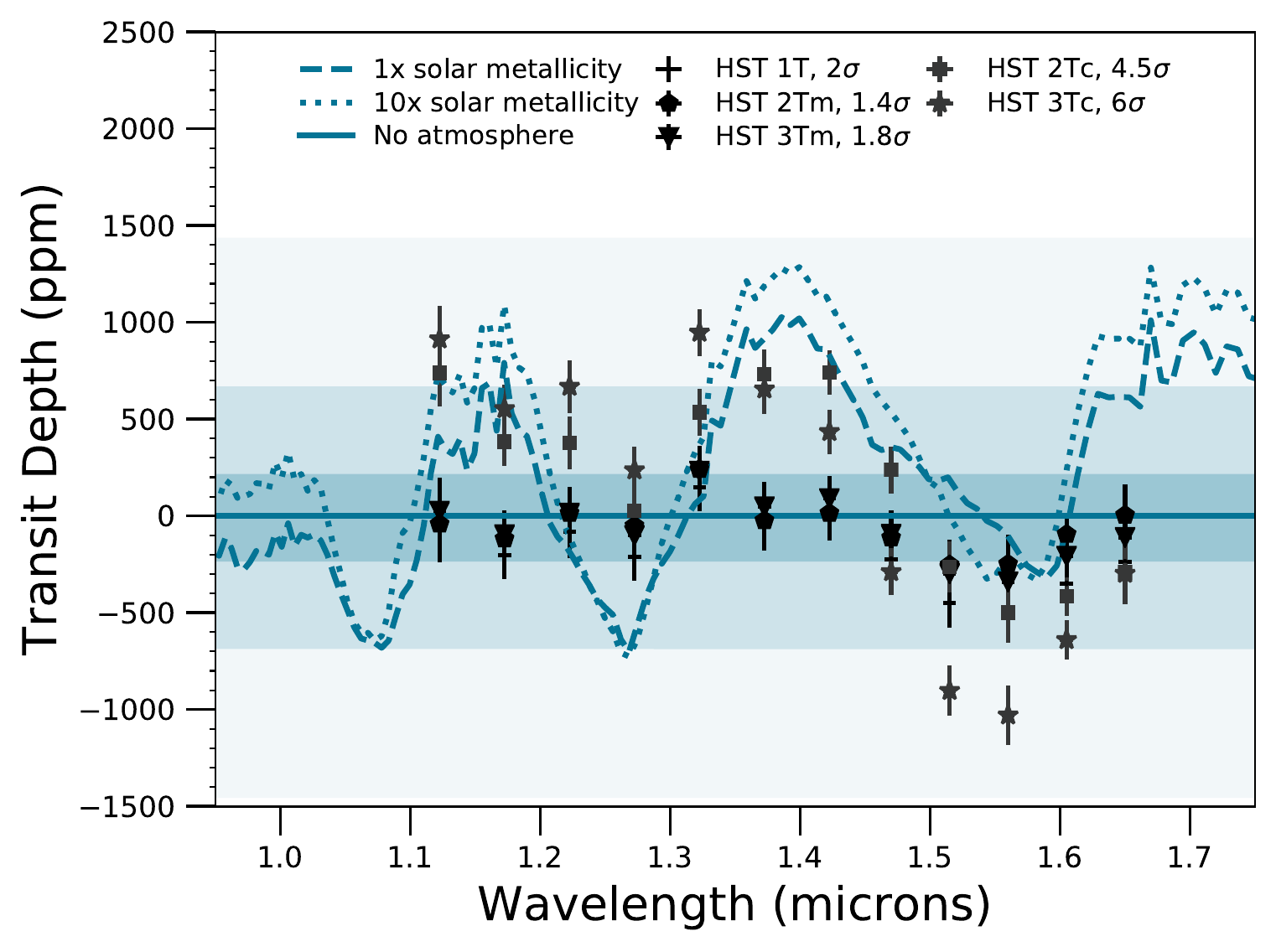}
    \includegraphics[width=0.9\linewidth]{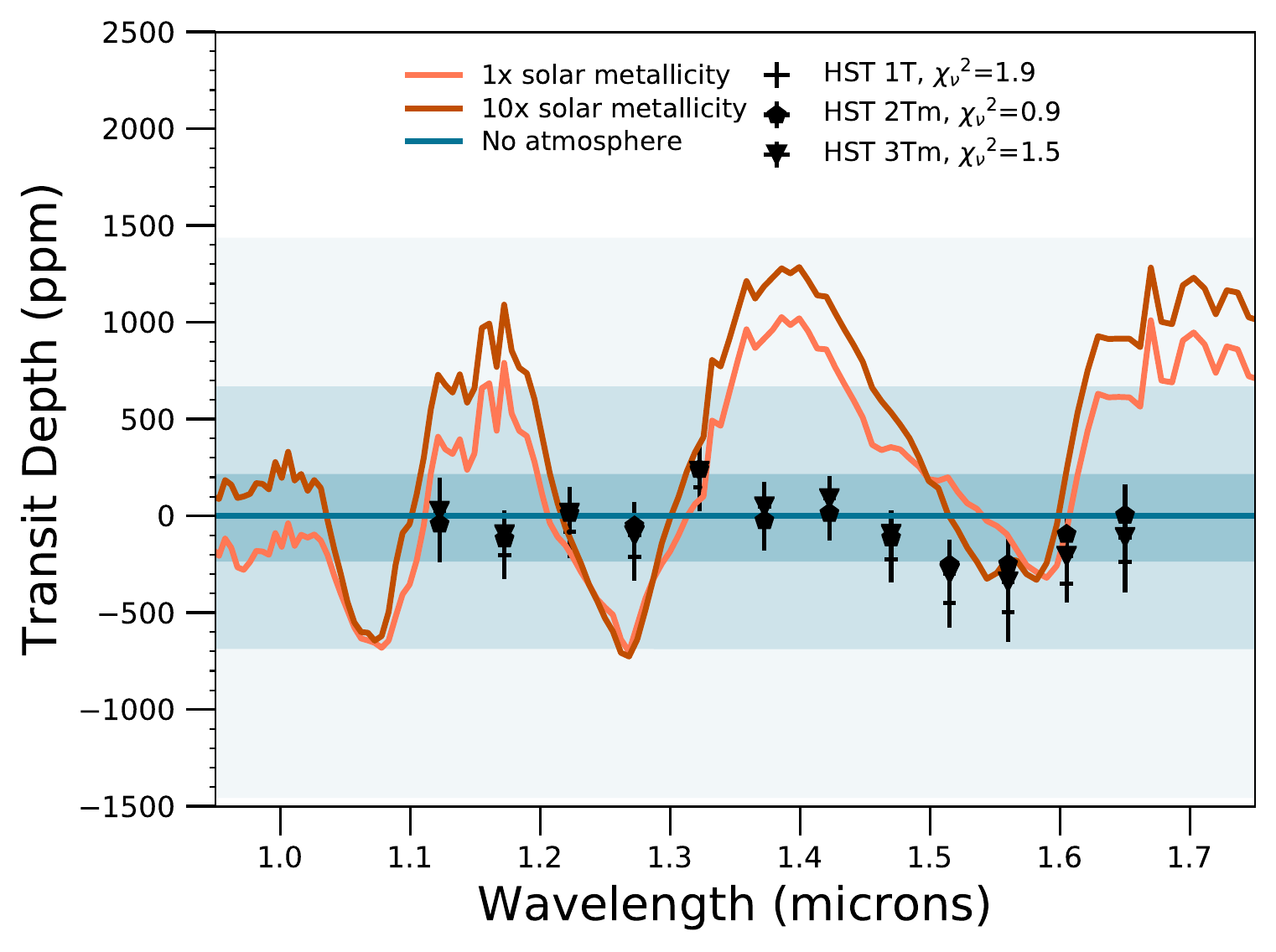}
    \caption{Our considered model atmospheres produced using \texttt{CHIMERA} compared to the five potential planetary spectrum scenarios. TOP: In addition to the mean-subtracted models (solid, dashed, and dotted lines), we include 1$\sigma$, 2$\sigma$, and 3$\sigma$ errors of the flat ``no atmosphere'' model as blue shaded regions. The confidence to which the solar metallicity scenario can be ruled out is reported after each scenario. The cold spot scenarios, 3Tc and 2Tc, can be clearly ruled out at high confidence. Of the remaining models and scenarios, no model can be confidently excluded at $\geq$3$\sigma$. However, a featureless infrared spectrum (solid line) either due to an airless body or an atmosphere without infrared features (e.g., a molecular nitrogen atmosphere) is statistically preferred over a 1 times solar metallicity atmosphere (dashed line) or a ten times solar, water- and methane-rich atmosphere (dotted line). BOTTOM:  \modif{Scenarios with the best goodness of fit} (1T, 2Tm, 3Tm) against the three model atmospheres (1$\times$ solar, light orange; 10$\times$ solar, dark orange; no atmosphere, blue). We report the \modif{$\chi_{\nu}^2$} of the flat line after the \modif{three} scenarios in the legend.}
    \label{fig:modelatmos}
\end{figure}

In \autoref{fig:modelatmos}, we show the results of our three modeled atmospheres for TRAPPIST-1 h compared to all five considered transit spectrum scenarios. In our figures, we show the data and models normalized by the mean of the flat model for better comparison. For scenarios 2Tc and 3Tc, we can rule out a solar metallicity H/He-dominated atmosphere at 4.5$\sigma$ and 6$\sigma$, respectively, \modif{where these $\sigma$ values are calculated from the critical values of our $\chi_{\nu}^2$ for the data and the models}. The water- and methane-rich atmosphere model excludes the 2Tc and 3Tc scenarios to greater than 5$\sigma$. While we cannot rule out the flat line model against the 2Tc/3Tc scenarios at $\geq$3$\sigma$ with the HST data alone, we can reject these potential spectra with relative confidence using the extended wavelength coverage of Spitzer, K2, and ground-based data, discussed further below.  

The remaining transit depth scenarios---1T, 2Tm, and 3Tm---fit to better than 3$\sigma$ for each atmosphere, limiting our ability to confidently discount a particular transit depth scenario or atmospheric model. However, with the HST WF3C data, we can rule out the ten times solar, water- and methane-rich atmosphere to nearly 2.5$\sigma$ for each of the transit depth scenarios and the 1 times solar metallicity to nearly 2$\sigma$. However, in each case,  we minimize our $\chi^2$ with the N$_2$ atmosphereless or airless body, aerosol-laden model. For 1T, 2Tm, and 3Tm, these values are \modif{$\chi_{\nu}^2$} = 1.9, 0.9, and 1.5, respectively, \modif{where $\chi_{\nu}^2 = \chi^2 / \nu$ with $\nu = 12$. Given a random draw of data from the model with the same uncertainty and resolution, one would get a model that would fit the data better 97.05\%, 45.39\%, and 88.43\% of the time for the 1T, 2Tm, and 3Tm scenarios, respectively. This is the $\chi^2$ cumulative distribution function (CDF), where values between 16\% and 84\% represent the 1$\sigma$ residual distribution \citep{wilson2021}. We report the $\chi^2$ CDF along with our $\chi_{\nu}^2$ values throughout the rest of the analysis.}

\begin{table*}
  \centering
  \begin{tabular}{ccl}
    \hline\hline
    Photospheric configuration &  Plausible?  & Explanation \\
    \hline
    1T & Yes & Cannot be ruled out beyond 1.4$\sigma$ with any atmospheric model \\

    \hline
    
    2Tm & Yes & Cannot be ruled out beyond 1$\sigma$ with any atmospheric model \\
    2Tc & Unlikely & Can be ruled out beyond 3.5$\sigma$ with any H$_2$-rich model; \\
     & & Can be ruled out to $\sim$1$\sigma$ with flat model \\
    
    \hline
    
    3Tm & Yes & Cannot be ruled out beyond 1.3$\sigma$ with any atmospheric model \\
    3Tc & No & Can be ruled out beyond 2.5$\sigma$ with every atmospheric model \\
    3Th & No & Hot spot coverage too small for planet to transit

  \end{tabular}
  \caption{Summary of which stellar photospheric configurations remain plausible when including the interpretation of planetary atmospheric models.}
  \label{tab:config_planet}
\end{table*}

\subsection{Model results with full wavelength coverage}

\begin{figure*}[htbp!]
  \centering
  \includegraphics[width=0.6\linewidth]{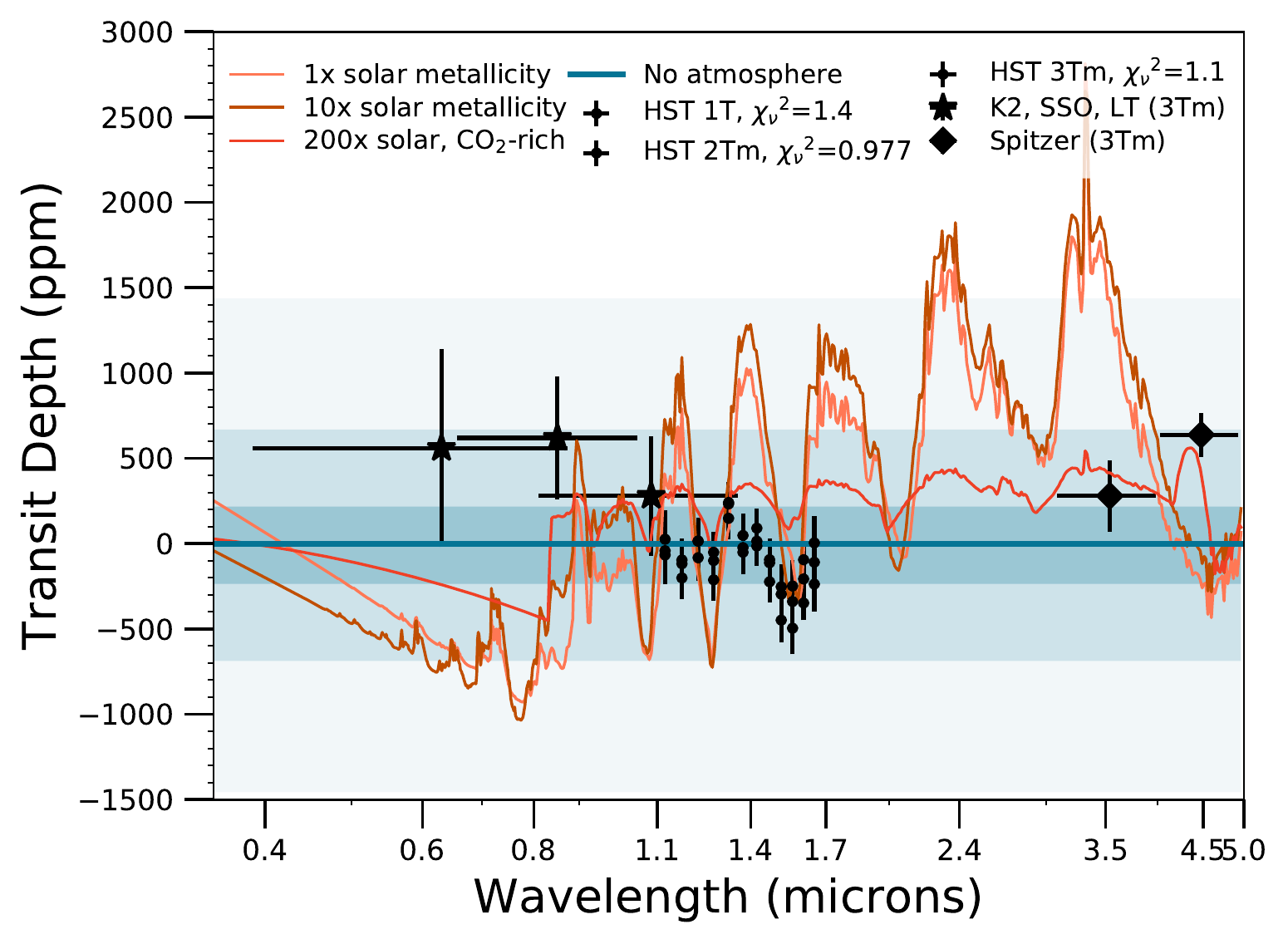}
  \caption{As in \autoref{fig:modelatmos} \modif{(bottom)} with the addition of K2/SSO/LT data points \citep{ducrot2018} and Spitzer IRAC Ch. 1 and Ch. 2 data \citep{ducrot2020}, shown here corrected for stellar contamination scenario 3Tm, the ``best-fit'' configuration. We also include our carbon-dioxide-rich atmospheric model to compare against the data, though in all scenarios for the stellar photosphere, we minimize the $\chi^2$ with the flat ``high mean molecular weight/no atmosphere'' model, which we report \modif{as $\chi_{\nu}^2$} after each transit depth scenario in the legend. We note that we have applied offsets to the K2, SSO, LT, and Spitzer data, as discussed in the text.}
  \label{fig:full_data_models}
\end{figure*}

In \autoref{fig:full_data_models}, we show the results of the previously modeled planetary atmospheres (1$\times$ and 10$\times$ solar metallicity and a ``flat'' model) in addition to a carbon dioxide-rich, 200 times solar metallicity atmosphere compared to an extended wavelength range from 0.8 $\mu$m to 4.5 $\mu$m of all existing transits of TRAPPIST-1 h. We include both the HST WFC3 G141 spectroscopic data analyzed here and additional photometric data from the space-based K2 campaign \citep{grimm2018}, the SPECULOOS-South Observatory (SSO) and the Liverpool Telescope (LT) on the ground \citep{ducrot2018}, and Spitzer/IRAC data at 3.6 $\mu$m and 4.5 $\mu$m \citep{ducrot2020}. We use the Spitzer transit depths reported by \citet{ducrot2020} in their Table 6, as these represent the depths of their global analysis. These data are therefore subject to fewer systematic errors between the two channels, in addition to being the preferred results by \citet{ducrot2020}. We then multiply these measured transit depths by the appropriate correction factor for each stellar configuration, as performed for the Spitzer 4.5 $\mu$m point in \citet{wakeford2019}.

Because of the potentially different systematic errors between the HST observations and those of the other telescope data \citep[e.g.,][]{dewit2016,dewit2018,ducrot2020}, we include various offsets for the transit depths for the Spitzer, K2, SSO, and LT data points. We allow these data to float within the range defined by their 1$\sigma$ errors and choose the offset that produces the minimum $\chi^2$ with the HST data \modif{after the appropriate correction factor is applied for each stellar photosphere scenario. Therefore, from the values given in \citealt{ducrot2020}, the K2 point was allowed to float $\pm$580 ppm, the LT point $\pm$350 ppm, the SSO point $\pm$360 ppm, and the Spitzer points $\pm$210 ppm}. We marginalize over the offset that favors the flattening of the K2, SSO, and LT data, as the large offsets between these measurements as reported by \citet{ducrot2018} require an unphysically large scattering slope in the NIR \citep[e.g.,][]{moran2018}.  As discussed below, the fact that relatively few transits were measured by K2, SSO, and LT (1, 3, and 1, respectively; \citealt{ducrot2018}) means that the associated error may also be higher than reported; however, these data do not ultimately guarantee the validity of or rule out any of the atmospheric models, given their already low precisions relative to the HST data. We treat the K2, SSO, and LT data as separate independent floats, but we treat the two Spitzer points together, using the larger error from the 3.6 $\mu$m data point as the allowable offset. 

We note that the preferred offset for the Spitzer points is the upper limit of how far we allow it to float, perhaps suggesting  an underestimation of the overall error assumed across our measurements. However, for consistency, we maintain the 1$\sigma$ error as the allowable offset range. In an effort to generate a potential atmosphere with molecular features that approach the transit depth of the 4.5 $\mu$m Spitzer point, we included the 200 times solar CO$_2$-rich atmosphere, where the presence of carbon monoxide produces a spectral feature in this range. We discuss \modiff{if such an atmosphere is physically realistic} in \autoref{implications}.

As stated above, the addition of the increased wavelength coverage allows us to fully discard the scenario 3Tc for the stellar photosphere. All atmospheric models are ruled out to well over 5$\sigma$ for 3Tc, except for the flat model, which is ruled out to 2.5$\sigma$, and the 200 times solar carbon-rich model, which is ruled out to 2.9$\sigma$. For scenario 2Tc, we can confidently exclude all but the flat and carbon-rich atmospheres, though we do increase our ability to reject the fit for 2Tc from less than 1$\sigma$ (0.7$\sigma$) with HST data alone to nearly 1$\sigma$ (0.9$\sigma$) with the full wavelength coverage. While not fully ruled out, we nevertheless choose not to include scenario 2Tc in \autoref{fig:full_data_models} as it is the least \deleted{statistically likely} \modif{preferred} of the remaining stellar photospheric scenarios. 

For the remaining stellar photospheres (1T, 2Tm, and 3Tm), the addition of the wider wavelength coverage does not allow us to rule out any atmospheric scenario with high confidence. For the one times solar metallicity atmosphere, we can only rule out scenario 2Tm to 0.5$\sigma$ and scenarios 3Tm and 1T to 0.6$\sigma$ and 0.7$\sigma$, respectively. For the water-rich ten times solar atmosphere, we can rule out transit depth scenarios 1T, 2Tm, and 3Tm to between 1.1 and 1.4 $\sigma$. Both the 200 times solar atmosphere and the featureless ``flat''/N$_2$ atmosphere models provide the best minimized $\chi^2$ fits to the remaining data. However, the flat model is slightly preferred over the carbon-dioxide-rich model. For scenarios 1T, 2Tm, and 3Tm, the carbon-dioxide-rich atmosphere results in \modif{$\chi_{\nu}^2$} = 4.0, 3.3, and 3.7.\modif{Using the $\chi^2$ CDF allows us to rule these out entirely, with a better dataset being drawn from the model 100\% of the time \citep{wilson2021}}. The flat model results in \modif{$\chi_{\nu}^2$} = 1.4, 1.0, and 1.1, with 17 degrees of freedom \modif{($\chi^2$ CDF of 87.50\%, 54.56\%, and 65.40\%) compared to scenarios 1T, 2Tm, and 3Tm}. As in \citet{wakeford2019}, this suggests the best fitting or  ``preferred'' stellar photospheric model is 3Tm, though we stress that we cannot actually constrain the stellar photosphere to high confidence with these existing measurements.

\begin{table*}
  \centering
  \begin{tabular}{clll}
    \hline\hline
    Planetary atmosphere & Description & Plausible? & Explanation \\
    \hline
    1$\times$ solar & Chemical equilibrium  & \modif{Strongly} unlikely & Poor fit to data, though not ruled out to 1$\sigma$\\
    & Rayleigh scattering & & \\
    & H$_2$, He, H$_2$O, CH$_4$, CO,  &  &  \\
    & CO$_2$, NH$_3$, N$_2$ & & \\
    \hline
    10$\times$ solar & Chemical equilibrium & \modif{Strongly} unlikely & Poor fit to data, but only ruled out to 1$\sigma$ \\
    & Rayleigh scattering & & \\
    & H$_2$, H$_2$O, CH$_4$ &  &  \\
    \hline
    200$\times$ solar & Chemical equilibrium & \modif{Unlikely} & Fits data with \modif{$\chi_{\nu}^2$} of $\sim$4 or better \\
    & Rayleigh scattering & & \\
    & CH$_4$, CO, CO$_2$ &  &  \\
    \hline
    No atmosphere/ & No Rayleigh scattering & Yes & Fits data with \modif{$\chi_{\nu}^2$} of $\sim$1.4 or better  \\
    Flat model/ & N$_2$ &  & \\
    Heavily aerosol-laden & & & \\
  \end{tabular}
  \caption{Summary of atmospheric models considered and their plausibility given the data}
  \label{tab:planetatmos}
\end{table*}

\subsection{Implications of planetary atmospheric models} \label{implications}

In constructing our planetary atmospheric models, we in part sought to maximize the potential differences between models to see if we could tell them apart based on the HST and additional wavelength data. While we cannot do so at high statistical significance, we ultimately determine that all evidence points to TRAPPIST-1 h having a high-mean-molecular-weight atmosphere ($\geq$1000$\times$ solar metallicity), a highly opaque aerosol layer, or no atmosphere, which cannot be distinguished with the current data precision.

The remaining models we show in \autoref{fig:modelatmos} and \autoref{fig:full_data_models} vary in terms of being physically realistic, which we explore here. The 1$\times$ and 10$\times$ solar metallicity model atmospheres are hydrogen-dominated. Such light atmospheres are most prone to atmospheric escape, though the location of TRAPPIST-1 h, that is,  farthest from the star, in principle means it is also most likely to have retained such an atmosphere against both hydrodynamic escape or via water photolysis, depending on its initial orbit and evolution \citep[e.g.,][]{bolmont2017,bourrier2017}. However, given the age of the TRAPPIST-1 system ($\sim$7 Gyr; \citealt{burgasser2017}), it is unlikely even for TRAPPIST-1 h to have retained a hydrogen-dominated atmosphere up until the present day, as it would lose the necessary hydrogen envelope to fit the mass--radius relationship given a terrestrial core in less than 100 Myr \citep{turbet2020}. Furthermore, \citet{turbet2020} argue that, given the total H-containing volatiles that any of the planets can accrete \citep{HoriandOgihara2020}, any individual planet that accreted a larger  fraction would result in a markedly different density from the others, but instead very similar densities are observed for all the planets in the TRAPPIST-1 system \citep{grimm2018,agol21}. Therefore, given our poor (though not statistically excluded) fits to the data with the hydrogen-dominated atmospheric models, combined with these theoretical concerns, we can conclude that TRAPPIST-1 h is highly unlikely to have a hydrogen-dominated atmosphere.

Given the size of the planet, which is between that of Mars and Earth, as well as its equilibrium temperature on the colder edge of that of Mars, we considered a 200 times solar metallicity, CO$_2$-rich atmosphere. However, at this cold a temperature ($\sim$170 K) and metallicity, both water and carbon dioxide are near the point where they can condense out of the atmosphere as crystalline ice clouds, or are vulnerable to atmospheric collapse entirely \citep{turbet2018}. Only very thick CO$_2$ atmospheres would be stable against collapse \citep{lincowski2018,turbet2018}, unless significant inventories of H$_2$ or CH$_4$ were present \citep{ramirezandkaltenegger2017,turbet2020mars}. This is similar to our naive 200 times solar CO$_2$-rich atmosphere. To demonstrate the unlikeliness of these CO$_2$ scenarios fitting the existing HST, Spitzer, and short-wavelength data, we calculate the absolute difference in transit depth between the Ch. 1 and Ch. 2 Spitzer points, which is approximately 350 ppm. For a pure CO$_2$ atmosphere (44 $\mu$), this corresponds to 50 scale heights (H), which is an unrealistically large extent for the planet. For our 200 times solar atmosphere with a mean molecular weight of $\sim$5 $\mu$, the difference in transit depths is 5H, which while large is not beyond the realm of possibility from theory \citep[e.g.,][]{millerricci2009}. Still, we cannot compellingly explain the Spitzer 4.5 $\mu$m with plausible models, despite not being able to rule them out statistically with the current data precision. Instead, a very high mean molecular weight atmosphere ($\geq$ 1000$\times$ solar metallicity) or no atmosphere at all remain the best explanations of the existing data, though we cannot tell these potential atmospheres (or lack thereof)  apart  at present.

Future observations are necessary to provide the higher precision and wavelength coverage needed to truly understand the existence and contents of an atmosphere around TRAPPIST-1 h. \modif{Furthermore, such higher quality data and wavelength coverage would allow future retrieval studies to place stronger quantitative constraints on the planet (e.g., \citealt{barstow2020})}. As part of General Observer Cycle 1, the JWST NIRSpec/PRISM will observe three transits of the planet from 0.6 to 5.3$\mu$m (JWST GO 1981; PIs Stevenson and Lustig-Yaeger), which will negate the need for offsets such as those calculated here, and will provide sufficient precision to distinguish between various high-metallicity atmospheres containing water, carbon dioxide, nitrogen, carbon monoxide, and methane. These observations, \modif{along with the atmospheric retrievals they enable}, can also offer further insight into the potential condensate clouds of TRAPPIST-1 h, constraints on which are beyond the precision of the existing HST, K2, SSO, LT, or Spitzer data.

\subsection{Planetary model summary}

In summary, we find that we can make several determinations about the nature of the star TRAPPIST-1 and the planet TRAPPIST-1 h using the combination of atmospheric models and potential stellar photospheres. We can fully discount photospheric configuration 3Th and 3Tc and tentatively discount scenario 2Tc. The remaining scenarios -- 1T, 2Tm, and 3Tm -- all produce reasonable transmission spectra to which we can fit atmospheric models, as summarized in \autoref{tab:config_planet}. For the planetary atmosphere, we can draw no strongly statistically significant conclusions based on the data; however, we do find some atmospheric models fit better than others for the allowable stellar configurations. While H$_2$-rich atmospheres cannot be ruled out to greater than 3$\sigma$, they show much poorer fits to the existing data from HST, even with the expanded wavelength coverage offered by K2, SSO, LT, and Spitzer. In all cases, a flat model representing either no atmosphere or a very high mean molecular weight atmosphere ($\geq$ 1000$\times$ solar metallicity) provide the best fit $\chi^2$s. These findings are also summarized in \autoref{tab:planetatmos}.

\section{Conclusion}\label{conclusion}
We present an analysis of the infrared transmission spectrum of TRAPPIST-1h, from 1.12 to 1.65 $\mu m$, obtained with the HST Wide Field Camera 3 and using the G141 grism. The spectra were extracted from the raw images with a pipeline based on the \texttt{prose} framework, and the resulting spectroscopic light curves modeled within a Bayesian framework, with a systematic error model selected through the minimization of the AIC. In order to disentangle the planetary spectrum from the stellar one (addressing the so-called stellar contamination effect), we modeled the median out-of-transit spectrum of TRAPPIST-1 against a multi-temperature combination of the PHOENIX BT-Settl theoretical models, following the approach of \cite{wakeford2019}. While the retrieved transmission spectra from each individual visit vary, none of our single-, two-, or three-component photospheric models (each component having a different temperature) is able to explain this variability. Nevertheless, our three-component model suggests the possibility that TRAPPIST-1 surface may be covered by 14.9\% $\pm$ 0.1\%  of cold spots (2000K $\pm$ 100K) and by a very small fraction (185.2 $\pm$ 8.8 ppm) of hot spots (hotter than 5000K). This result is compatible with the results of \citet{morris2018}, explaining the varying amplitude of the photometric variability of TRAPPIST-1 observed between K2 and Spitzer. However, we find that none of our photosphere models, including a homogeneous photosphere, provide a good fit to the data. While it might come from the accuracy of the PHOENIX models being used, our approach requires constraints on the photospheric structure of TRAPPIST-1 in order to break the multiple degeneracies encountered when fitting multi-component models to the data. This prior knowledge is not yet available for ultra-cool dwarf stars, but will certainly be essential for the atmospheric characterizations to come. To this end, tools like the ensemble analysis methodology described in \cite{luger2021}, or transmission spectroscopy of active region occultations \citep{espinoza2019} offer promising avenues. \modif{Finally, we draw particular attention to the very simplistic nature of the active region modeling used in \autoref{stellar-contamination},  which is foreseen to be improved by further study of ultra-cool dwarf star photospheres as these objects are particularly valuable to the study of small exoplanets.}
\bigskip\\
With the potential stellar photospheric scenarios in hand, we modeled a number of different planetary atmospheres using the forward model \texttt{CHIMERA}, including various H$_2$-dominated atmospheres, a carbon-dioxide-rich atmosphere, and a featureless ``flat'' line model. While the data quality combined with the cold, small (likely rocky) nature of the planet are not sufficient to exclude any atmospheric scenario to high statistical significance, we conclude that TRAPPIST-1 h is highly unlikely to possess a hydrogen-dominated atmosphere \added{(a conclusion independently reached by \citealt{Gressier2022})}. The likeliest scenario for this planet is that it possesses a very high mean molecular weight ($\geq$1000$\times$ solar metallicity) atmosphere, is enshrouded by an opaque aerosol layer, or is devoid of atmosphere entirely. Determination of the true nature of the atmosphere of TRAPPIST-1 h, or lack thereof, awaits upcoming measurements by JWST. 

\section{Acknowledgment}
This research made use of \textsf{exoplanet} \citep{exoplanet} and its dependencies \citep{exoplanet:agol20, exoplanet:arviz, exoplanet:astropy13, exoplanet:astropy18, exoplanet:exoplanet, exoplanet:luger18, exoplanet:pymc3, exoplanet:theano}. We acknowledge support from the BELSPO program BRAIN-be 2.0 (Belgian Research Action through Interdisciplinary Networks) contract B2/212/B1/PORTAL. \modif{This publication benefits from the support of the French Community of Belgium in the context of the FRIA Doctoral Grant awarded to L.J.G.} MG is F.R.S.-FNRS Senior Research Associate. B.V.R. thanks the Heising-Simons Foundation for support. S.E.M. acknowledges support from NASA Earth and Space Science Fellowship grant 80NSSC18K1109. \modif{Finally, this work is based on observations made with the NASA/ESA Hubble Space Telescope that were obtained at the Space Telescope Science Institute, which is operated by the Association of Universities for Research in Astronomy, Inc. These observations are associated with programs and grants under GO-15304 (PI. J. deWit)}

\bibliographystyle{aa}
\bibliography{bib}

\appendix
\section{Complementary figures}
\subsection{Spectra extraction and modeling comparison}

\begin{figure*}
  \centering
  \includegraphics[width=0.9\linewidth]{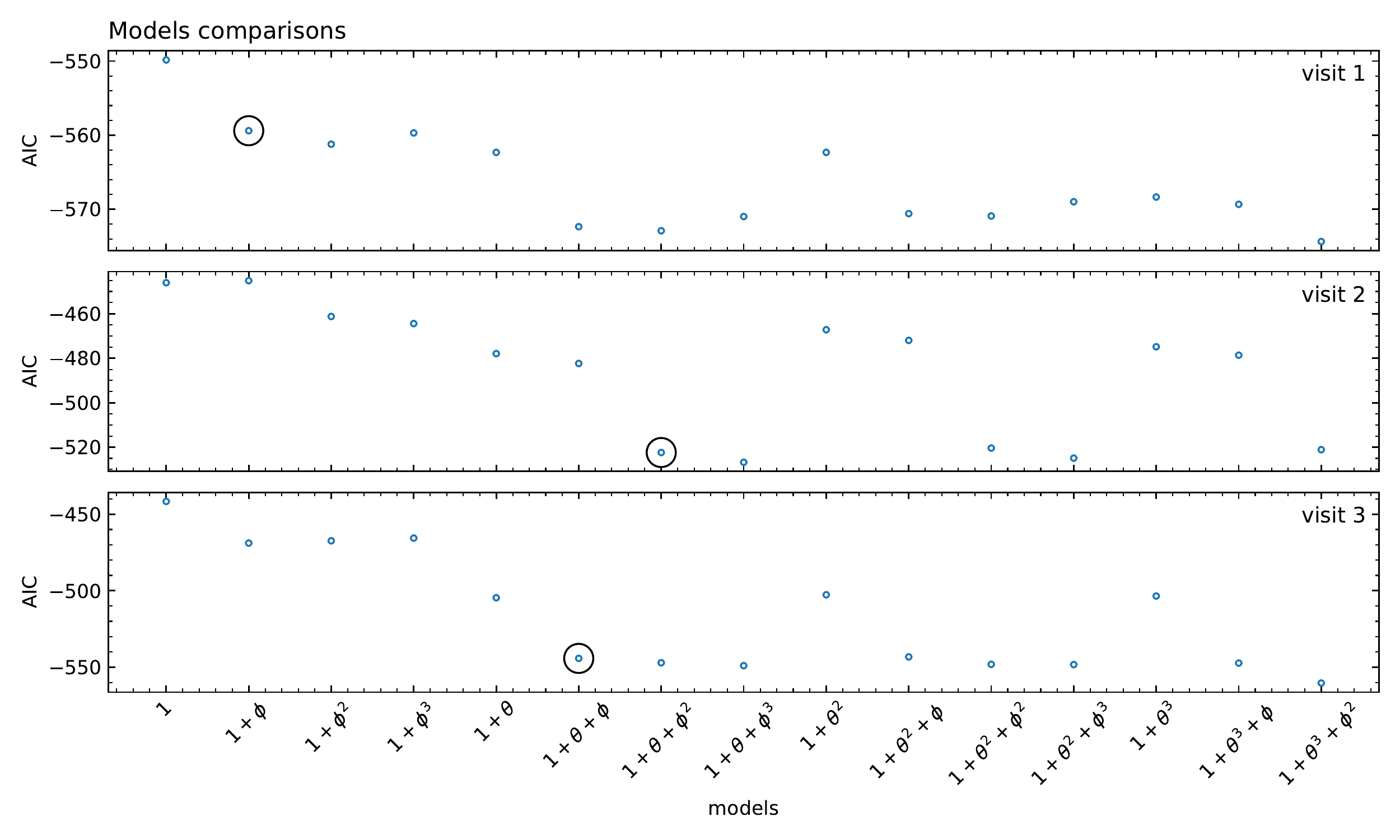}
  \caption{Model comparison for visits \modif{1, 2, and 3}. The x-axis represents the systematic error model where $1$ designates a constant, $\theta^n$ a polynomial order $n$ of $\theta$, and $\phi^m$ a polynomial order $m$ of $\phi$. The {best} model is circled. For one model to be considered {better} than another, we require that the AIC difference between it and the others be greater than \modif{20}.}
  \label{fig:aics}
\end{figure*}

\begin{figure*}
  \centering
  \includegraphics[width=\linewidth]{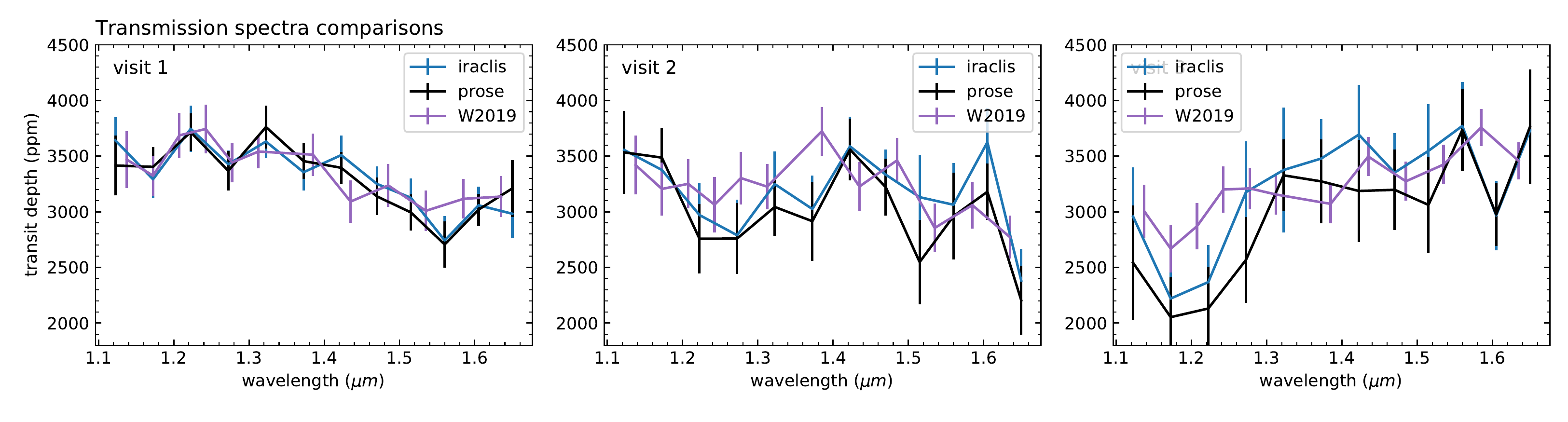}
  \caption{Comparison of the transmission spectra of our three HST visits obtained with \textsf{iraclis}, \textsf{prose,} and the method from \cite{wakeford2019}. The larger discrepancy of visit 3 for the method of \cite{wakeford2019} (especially between 1.1 and 1.3 $\micron$) is due to the partial masking of the first orbit, whereas the other two analyses discard it.}
  \label{fig:iracomp}
\end{figure*}

\begin{figure}
  \centering
  \includegraphics[width=\linewidth]{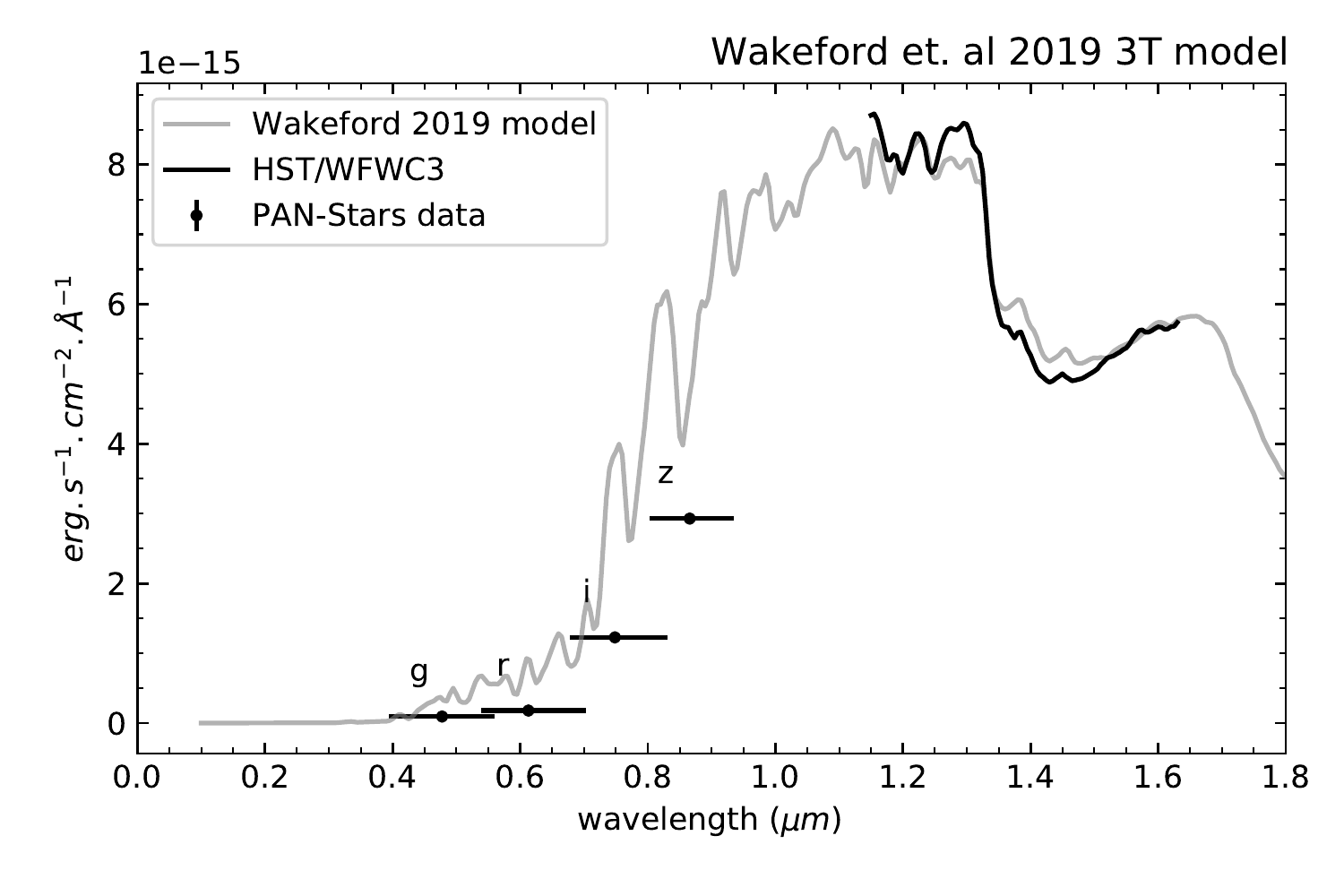}
  \caption{\citealt{wakeford2019} three-component model against our HST data and Pan-STARSS g, r, i, and z wide-band photometry}
  \label{fig:hw}
\end{figure}


\begin{figure}[htbp!]
  \centering
  \includegraphics[width=\linewidth]{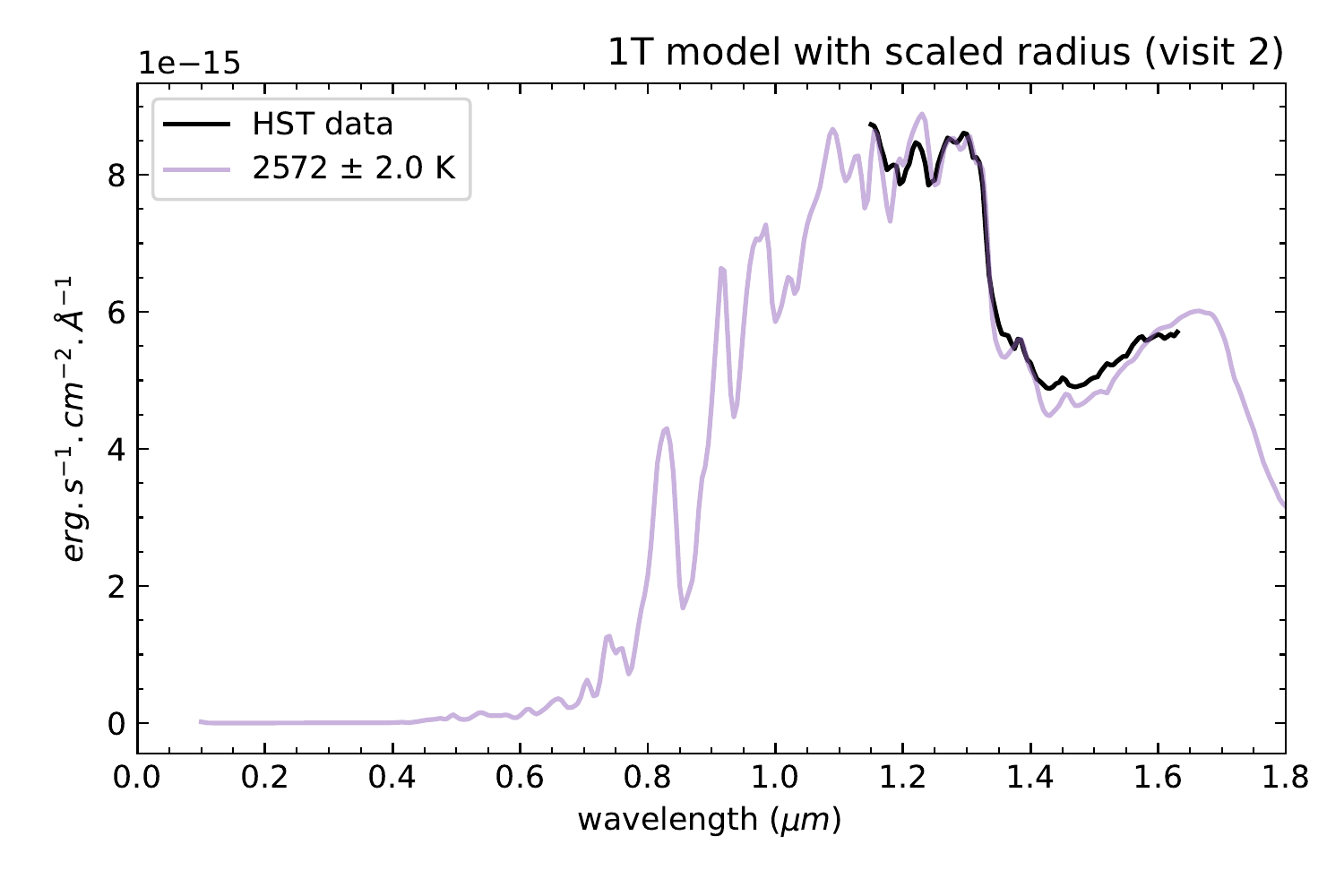}
  \caption{Visit 2 single-component photospheric model, allowing for a scaled radius (as in \citealt{wakeford2019})}
  \label{fig:scaled_oot_fit}
\end{figure}

\begin{figure*}[htbp!]
  \centering
  \includegraphics[width=0.65\linewidth]{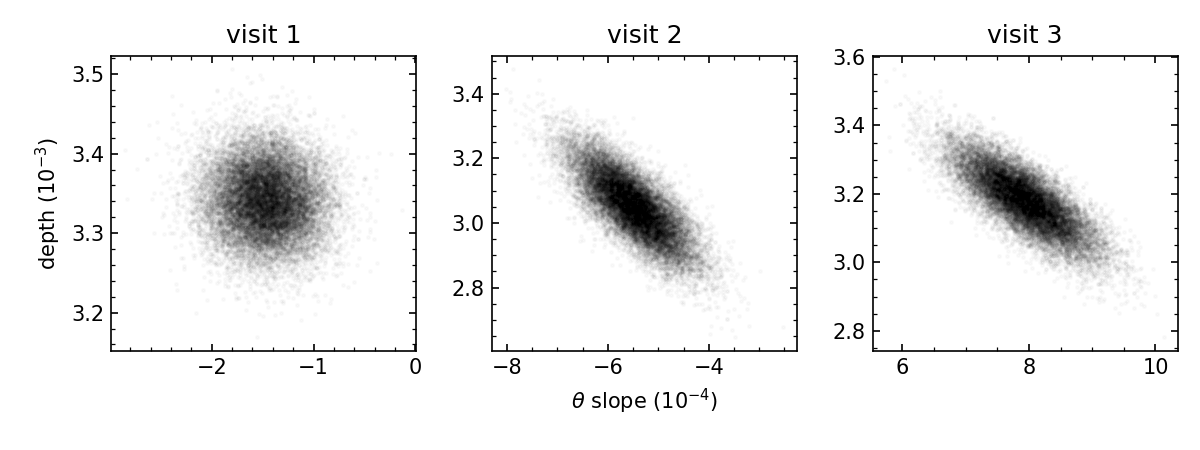}
  \caption{\modif{Correlation between the slope in time and transit depth of the systematic error model. This figure was produced by sampling the likelihood distribution of our model against its parameters using a Markov Chain Mont Carlo on the white-light curves. It shows that, apart from visit 1 where the transit mid-time is observed, the transit depth is correlated with the slope in time $\theta$ used to model \modiff{systematics over the duration of a visit} (a shallower slope is compensated by a deeper transit). As only the ingress and egress are observed in visits 2 and 3, we assume that the spectrum-to-spectrum differences in our study are partially due to this effect. However, as the visits-combined analysis contains a complete transit (in phase), the slope of the  systematic error model should be less correlated with the transit depth, as in visit 1, which is confirmed by the similarity observed between visit 1 and the global spectrum (\autoref{fig:transp})}}
  \label{fig:corre}
\end{figure*}

\subsection{Out-of-transit spectra inference}\label{oot-corner}
\begin{figure}[htbp!]
  \centering
  \includegraphics[width=0.4\linewidth]{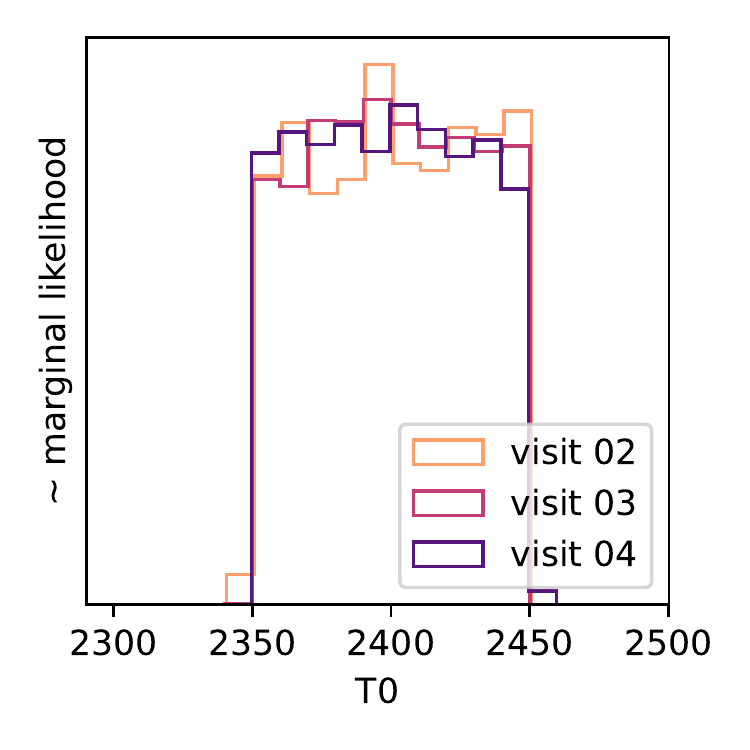}
  \caption{Corner plot of the parameters inferred in the 1T model. \modif{$T_0$ denotes the temperature of the cooler component})}
\end{figure}
\begin{figure}[htbp!]
  \centering
  \includegraphics[width=\linewidth]{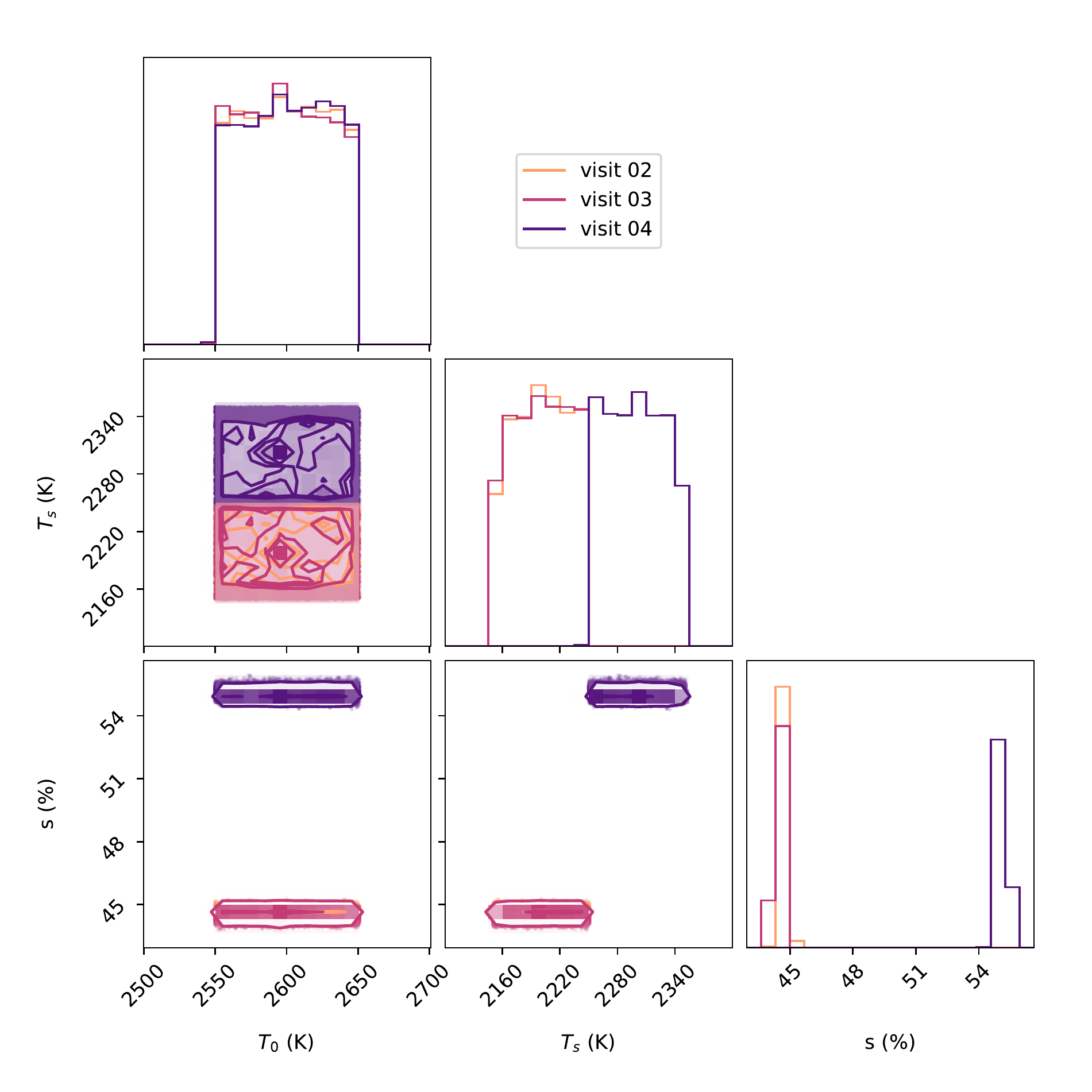}
  \caption{Corner plot of the parameters inferred in the 2T model. \modif{Here $T_0$ denotes the temperature of the quiescent component and $T_s$ the temperature of the cooler component, with $s$ its covering fraction}. Visit \modif{3} is better modeled with a cooler component at $\sim$2300K instead of $\sim$2200K for visit \modif{1 and 2}. We assume that this is likely due to the coarse grid of temperatures of our model, as the likelihood distribution seems to have a single maximum (see \autoref{fig:2T-ll}).}
\end{figure}
\begin{figure}[htbp!]
  \centering
  \includegraphics[width=0.9\linewidth]{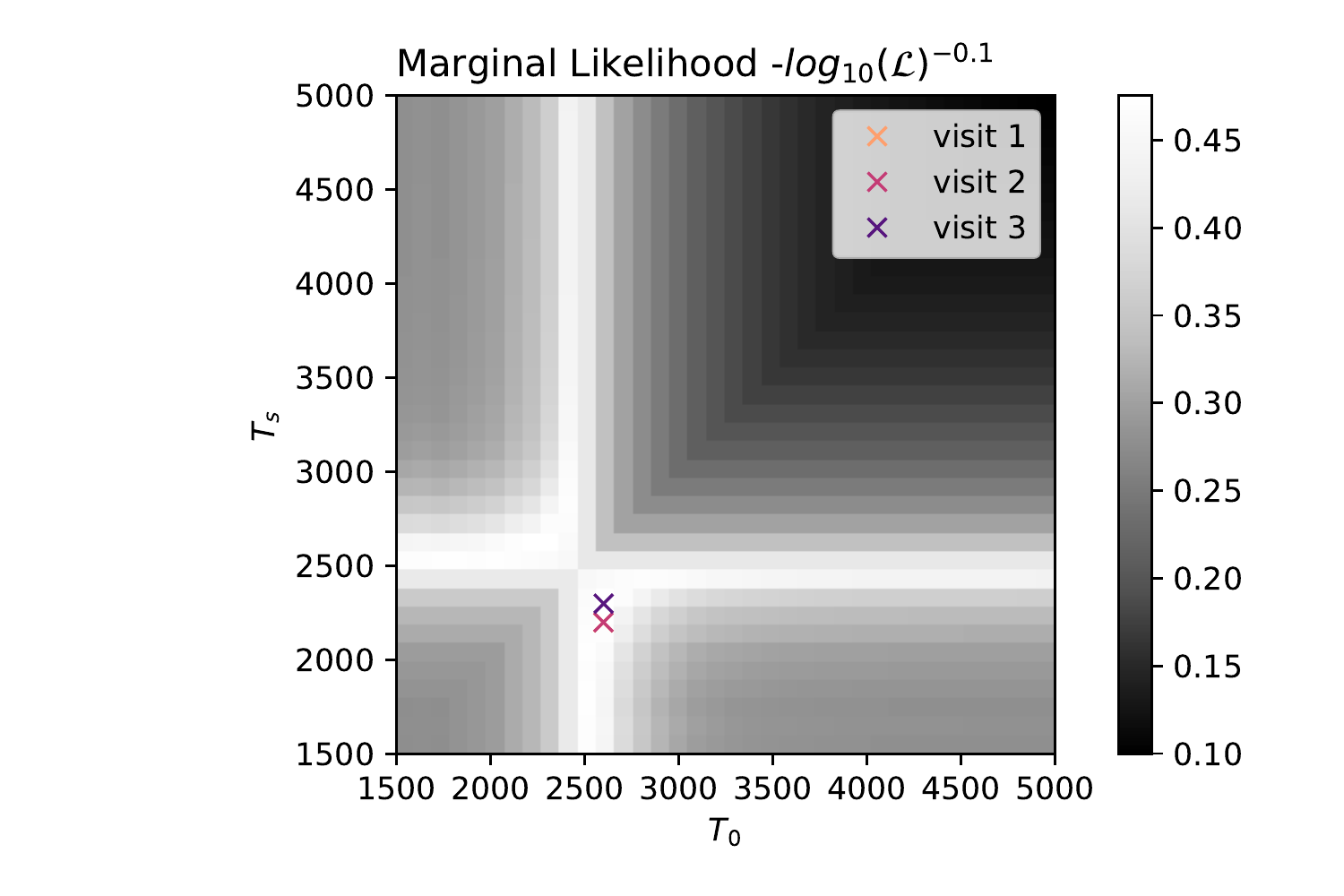}
  \caption{2T model likelihood sampled on the temperature grids described in \autoref{tab:grids_details} and marginalized over the photospheric heterogeneity covering fractions. The symmetry of this distribution is simply due to the symmetry of our two-component model, with covering fractions $f$ and $1-f$.}
  \label{fig:2T-ll}
\end{figure}

\begin{figure*}[htbp!]
  \centering
  \includegraphics[width=0.8\linewidth]{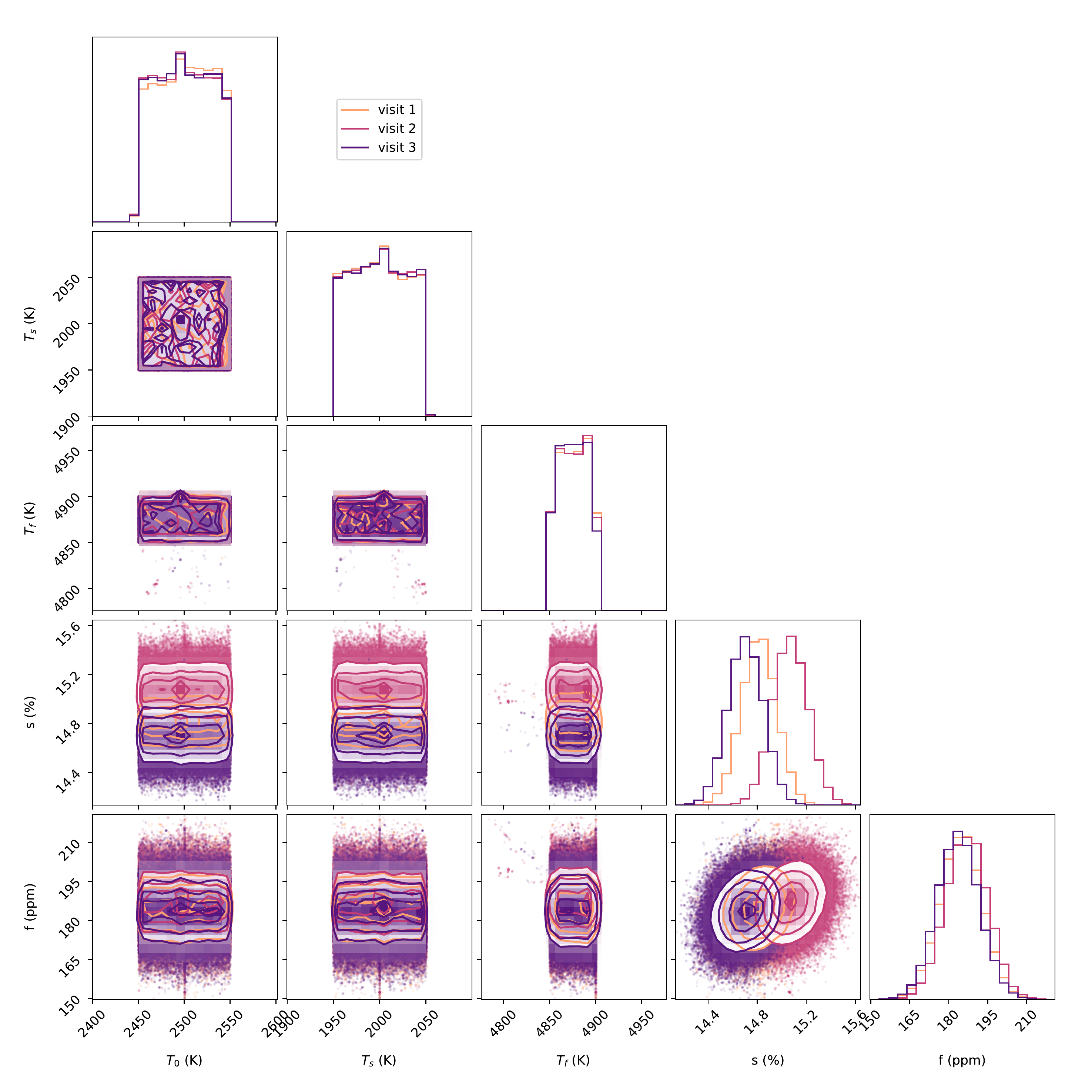}
  \caption{Corner plot of the parameters inferred in the 3T model. \modif{Here $T_0$ denotes the temperature of the quiescent component, $T_s$ the temperature of the cooler component with $s$ its covering fraction and $T_f$ the temperature of the hotter component with $f$ its covering fraction.}}
\end{figure*}

\begin{figure*}[htbp!]
  \centering
  \includegraphics[width=0.6\linewidth]{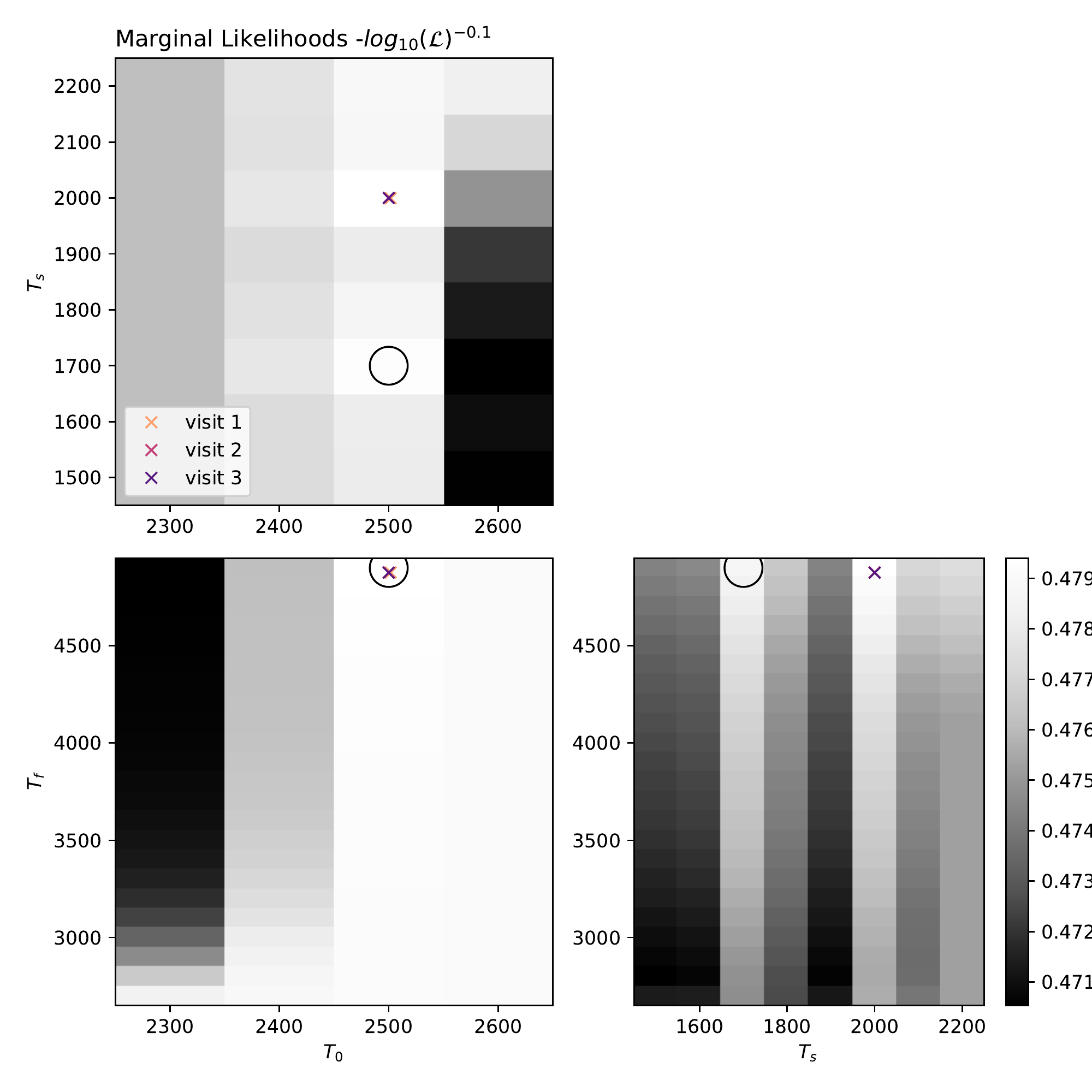}
  \caption{3T model likelihood sampled on the temperature grids described in \autoref{tab:grids_details} and marginalized over the covering fractions of the photospheric heterogeneities, as well as the temperature of each component. For each visit, colored crosses correspond to the global maximum likelihood parameters. The black circle shows a local maximum of the likelihood distribution, highlighting the bimodal nature of the model.}
  \label{fig:3T-ll}
\end{figure*}

\end{document}